\documentclass[11pt, a4paper]{article}    %

\usepackage[T1]{fontenc} 						%
\usepackage[utf8]{inputenc} 				%
\usepackage[english]{babel}
\usepackage{cancel}
\usepackage{amsfonts,amssymb,amsmath,mathtools}
\usepackage[hidelinks]{hyperref}
\usepackage{graphicx}
\usepackage[inline,shortlabels]{enumitem}
\usepackage{soul}
\usepackage{algorithm,tabularx}
\usepackage[noend]{algpseudocode}
\usepackage{cite}
\usepackage{tcolorbox}

\newcommand{\R}{\mathbb{R}}

\newcommand{\N}{\mathbb{N}}

\newcommand{\T}{^\top}
\newcommand{\inv}{^{-1}}

\newcommand{\subj}{\tn{subj.~to}}

\DeclareMathOperator{\blkdiag}{blkdiag}
\DeclareMathOperator{\col}{col}
\DeclareMathOperator\rank{rank}
\DeclareMathOperator{\Tr}{Tr}

\newtheorem{theorem}{Theorem}[section]
\newtheorem{proposition}[theorem]{Proposition}

\newtheorem{lemma}[theorem]{Lemma}
\newtheorem{assumption}[theorem]{Assumption}
\newtheorem{remark}[theorem]{Remark}

\newtheorem{proof}[theorem]{Proof}

\newcommand\oprocendsymbol{\hbox{$\blacksquare$}}
\newcommand\oprocend{\relax\ifmmode\else\unskip\hfill\fi\oprocendsymbol}

\newcommand{\tn}[1]{{\textnormal{#1}}}

\newcommand{\cA}{\mathcal A}
\newcommand{\cB}{\mathcal B}
\newcommand{\cC}{\mathcal C}

\newcommand{\cK}{\mathcal K}

\newcommand{\cS}{\mathcal S}

\newcommand{\cW}{\mathcal W}
\newcommand{\cX}{\mathcal X}

\newcommand{\until}[1]{\{1,\ldots,#1\}} 

\graphicspath{{figs/}}

\newcommand{\rmv}{\mathrm{v}}

\newcommand{\norm}[1]{\left \|#1 \right \|}
\newcommand{\snorm}[1]{ \|#1 \|}

\newcommand{\rz}{\mathrm{z}}
\newcommand{\ry}{\mathrm{y}}

\newcommand{\ru}{\mathrm{u}}

\newcommand{\rd}{\mathrm{d}}

\newcommand{\uu}{\ru}
\newcommand{\z}{\rz}
\newcommand{\y}{\ry}
\newcommand{\dd}{\rd}

\newcommand{\rg}{C} %
\newcommand{\ms}{\y} %
\newcommand{\es}{\theta} %
\newcommand{\pr}{\theta_\star} %
\newcommand{\rr}{S} %

\newcommand{\ud}{_}%
\newcommand{\iter}{t}
\newcommand{\initer}{\tau}
\newcommand{\iterp}{{t+1}}
\newcommand{\timeid}{\iter}
\newcommand{\timeidp}{{\iter + 1}}

\newcommand{\prt}{\theta\ud{\star,\iter}}

\newcommand{\mst}{\ms\ud{\iter}}

\newcommand{\est}{\es\ud{\iter}}
\newcommand{\estp}{\es\ud{\iterp}}
\newcommand{\rrt}{\rr\ud{\iter}}
\newcommand{\rrtp}{\rr\ud{\iterp}}

\newcommand{\trr}{\tilde{\rr}}

\newcommand{\ths}{\tilde{\hs}}

\newcommand{\ut}{\uu\ud{\iter}}
\newcommand{\utime}{\uu\ud{\timeid}}

\newcommand{\tes}{\tilde{\es}}

\newcommand{\av}{^{ \textsc{\tiny av}}}

\newcommand{\tesa}{\tilde{\es}\av}
\newcommand{\tesat}{\tesa\ud{\iter}}

\newcommand{\tesap}{\tesa\ud{+}}

\newcommand{\n}{n}

\newcommand{\rx}{\mathrm{\x}}
\newcommand{\x}{\mathrm{x}}

\newcommand{\xt}{\rx\ud\iter}
\newcommand{\xtp}{\rx\ud{\iterp}}

\newcommand{\xtime}{\rx\ud\timeid}
\newcommand{\xtimep}{\rx\ud{\timeidp}}

\newcommand{\dr}{\dd}
\newcommand{\drt}{\dr\ud\iter}

\newcommand{\zt}{\z\ud\iter}
\newcommand{\ztp}{\z\ud{\iterp}}

\newcommand{\rw}{\mathrm{w}}
\newcommand{\w}{\rw}

\newcommand{\wt}{\w\ud{\iter}}
\newcommand{\wtp}{\w\ud{\iterp}}

\newcommand{\g}{K} %

\newcommand{\gt}{\g\ud{\iter}}
\newcommand{\gtp}{\g\ud{\iterp}}

\newcommand{\tg}{\tilde{\g}} %

\newcommand{\tga}{\tg\av} %
\newcommand{\tgat}{\tga\ud{\iter}}

\newcommand{\tgap}{\tga\ud{+}} %

\newcommand{\Vav}{V}

\newcommand{\gstar}{\g_\star}
\newcommand{\gstart}{\g_{\star,\iter}}

\newcommand{\G}{G}
\newcommand{\J}{J}

\newcommand{\za}{\z\av}
\newcommand{\zat}{\za\ud\iter}
\newcommand{\zatp}{\za\ud\iterp}

\newcommand{\tza}{\tz\av}
\newcommand{\tzat}{\tza\ud\iter}
\newcommand{\tzatp}{\tza\ud\iterp}

\newcommand{\vc}[1]{\mathtt{vec}\left (#1 \right )}
\newcommand{\svc}[1]{\mathtt{vec}\!\left (#1 \right )}
\newcommand{\ssvc}[1]{\mathtt{vec}\!\!\left (#1 \right )}
\newcommand{\unvc}[1]{\mathtt{unvec}\left (#1 \right )}

\newcommand{\dpr}{\R^{(n + m) \times n}}

\newcommand{\hs}{H} %
\newcommand{\hst}{\hs\ud{\iter}}
\newcommand{\hstp}{\hs\ud{\iterp}}

\newcommand{\lipp}{\beta}

\newcommand{\dimx}{{n}}
\newcommand{\dimu}{{m}}
\newcommand{\bE}{\mathbb{E}}

\newcommand{\gd}{\mu}

\newcommand{\At}{{A}\ud\iter}
\newcommand{\Bt}{{B}\ud\iter}
\newcommand{\As}{{A}\ud\star}
\newcommand{\Asv}{A^\iter_\star}
\newcommand{\Bs}{{B}\ud\star}
\newcommand{\Bsv}{B^\iter_\star}

\newcommand{\mav}{\kappa}

\newcommand{\M}{M}

\newcommand{\pix}{\Pi_{\x}}
\newcommand{\pih}{\Pi_{\hs}}
\newcommand{\pir}{\Pi_{\rr}}

\newcommand{\crr}{\mathcal{\rr}}
\newcommand{\chs}{\mathcal{\hs}}

\newcommand{\harm}{q}
\newcommand{\nw}{n_\w}
\newcommand{\EE}{E}

\newcommand{\sts}{\chi^{\text{\tiny ss}}} %

\newcommand{\halff}{\tfrac{1}{2}}

\newcommand{\Dt}{\cW\ud\iter}
\newcommand{\Ct}{\cC\ud\iter}

\newcommand{\hsss}{\hs^{\tn{ss}}}
\newcommand{\hssst}{\hsss\ud\iter}

\newcommand{\per}{\iter_{\w}}

\newcommand{\ssz}{\gamma}

\newcommand{\vv}{^{\text{vc}}}

\newcommand{\vhs}{\hs\vv}
\newcommand{\vhst}{\vhs\ud\iter}
\newcommand{\vhstp}{\vhs\ud\iterp}

\newcommand{\vrr}{\rr\vv}
\newcommand{\vrrt}{\vrr\ud\iter}
\newcommand{\vrrtp}{\vrr\ud\iterp}

\newcommand{\F}{F}

\newcommand{\tPE}{\mathbf{t}_{\dr}}

\newcommand{\Sw}{F}

\newcommand{\chit}{\chi\ud\iter}
\newcommand{\chitp}{\chi\ud\iterp}
\newcommand{\fun}{\phi}

\newcommand{\nc}{n_{\chi}}
\newcommand{\cAc}{\cA_{\g}}

\newcommand{\tc}{\tilde{\chi}} 

\newcommand{\tct}{\tc\ud\iter}

\newcommand{\tctp}{\tc\ud\iterp}

\newcommand{\nz}{{n_{\z}}}

\newcommand{\tz}{\tilde{\z}}

\newcommand{\tzt}{\tz\ud\iter}
\newcommand{\tztp}{\tz\ud\iterp}

\newcommand{\tr}{t^{\text{mid}}}
\newcommand{\var}{\sigma}

\newcommand{\opvecunvec}{\rmv}
\newcommand{\opH}{\opvecunvec_{H}}
\newcommand{\opS}{\opvecunvec_{S}}

\newcommand{\Vs}{\cS_{\text{\small ss}}}

\newcommand{\scalebd}{1}
\newcommand{\scalefig}{1}
\newcommand{\scalefigzoom}{1}

\def\algname/{\textsc{Relearn} LQR}

\newcommand{\rmax}{r^{\text{\tiny M}}}

\newcommand{\set}{Z}

\usepackage{soul}

\newcommand{\mathsout}[1]%
{\bgroup\mathchoice
  {\sbox0{$\displaystyle{#1}$}%
    \usebox0\hspace{-\wd0}%
    \rule[0.5\ht0-0.5\dp0-.5pt]{\wd0}{1pt}}%
  {\sbox0{$\textstyle{#1}$}%
    \usebox0\hspace{-\wd0}%
    \rule[0.5\ht0-0.5\dp0-.5pt]{\wd0}{1pt}}%
  {\sbox0{$\scriptstyle{#1}$}%
    \usebox0\hspace{-\wd0}%
    \rule[0.5\ht0-0.5\dp0-.5pt]{\wd0}{1pt}}%
  {\sbox0{$\scriptscriptstyle{#1}$}%
    \usebox0\hspace{-\wd0}%
    \rule[0.5\ht0-0.5\dp0-.5pt]{\wd0}{1pt}}%
	\egroup}
	
	\usepackage[textwidth=1.4cm,textsize=scriptsize]{todonotes}
\title{Stability-Certified On-Policy Data-Driven LQR\\via Recursive Learning and Policy Gradient}

\author{Lorenzo Sforni, Guido Carnevale,\\Ivano Notarnicola, Giuseppe Notarstefano\thanks{The authors are with the Department of Electrical, Electronic and Information Engineering, Alma Mater Studiorum - Universit\`{a} di Bologna, Bologna, 40136, Italy. (\texttt{name.surname@unibo.it}).
A preliminary short version of this paper is~\cite{sforni2023on}. 
The present article includes an improved comprehensive treatment, all the theoretical proofs, and a concrete application.
The corresponding author is L.~Sforni \texttt{lorenzo.sforni@unibo.it}.}}
	
\begin{document}
\maketitle
	
	\begin{abstract}
		In this paper, we investigate a data-driven framework to solve
		Linear Quadratic Regulator (LQR) problems when the dynamics
		is unknown, with the additional challenge of providing
		stability certificates for the overall learning and control scheme.
		Specifically, in the proposed on-policy learning framework,
		the control input is applied to the actual (unknown) linear
		system while iteratively optimized.
		We propose a learning and control procedure, termed
		\algname/, that combines a recursive least squares method
		with a direct policy search based on the gradient
		method. The resulting scheme is analyzed by modeling it as a
		feedback-interconnected nonlinear dynamical system.
		A Lyapunov-based approach, exploiting averaging and timescale separation theories for
		nonlinear systems, allows us to provide formal stability
		guarantees for the whole interconnected scheme.  The
		effectiveness of the proposed strategy is corroborated by
		numerical simulations, where \algname/ is deployed on an
		aircraft control problem, with both static and drifting
		parameters.
	\end{abstract}

\section{Introduction}

The massive availability of data in automation and robotics pushed the control community to revise the traditional model-based Optimal Control (OC) approaches toward learning-driven scenarios.
In this context, the control policy is iteratively updated without an explicit knowledge of the underlying dynamical system, relying solely on the collected data.
Hence, the key distinction between off-policy and on-policy methods arises from the interconnection between gathered data and the current policy.
Specifically, off-policy algorithms pursue a value iteration approach, and data are, in general, independent of the current policy.
Conversely, on-policy algorithms employ a policy iteration framework, evaluating the performance of the current policy using data generated under the same policy.
Since the early derivations of Reinforcement Learning (RL) methods for Linear Quadratic (LQ) regulation~\cite{bradtke1994adaptive}, there has been growing interest in data-driven solutions to infinite-horizon Linear Quadratic Regulator (LQR).
The recent survey~\cite{recht2019tour} investigates connections between OC and RL.

In the off-policy context, we find iterative methods inspired by the Kleinman algorithm~\cite{kleinman1968on}, involving either parameter identification or direct policy estimation~\cite{pang2021robust,lopez2023efficient, qin2014online, krauth2019finite,modares2016optimal, pang2018data}.  
The recent works~\cite{possieri2022q, bian2016value, ziemann2022policy} propose algorithms that do not assume the existence of stabilizing initial policies.
A model-free approach for discrete-time LQR based on RL is studied and developed in~\cite{kiumarsi2017h}. 
Off-policy approaches can be further distinguished between direct, where data are used directly in the policy design phase, and indirect approaches, where a preliminary identification step is performed.
Direct strategies often tackle the LQR problem by exploiting Persistently Exciting (PE) data together with semi-definite programming and Linear Matrix Inequalities (LMI) approaches, as introduced in~\cite{de2019formulas} and extended in~\cite{van2020data,rotulo2020data,rotulo2022online}. 
These LMI-based solutions also allowed for the design of control policies in case of noisy data, as explored in~\cite{de2021low,dorfler2023certainty}.
Direct approaches have been deployed to address also the design of robust controllers, e.g., in~\cite{berberich2020robust,van2020noisy}.
The recent survey~\cite{depersis2023learning} also includes an extension to nonlinear systems.
Instead, indirect approaches are explored~\cite{dean2019safely, mania2019certainty, ferizbegovic2019learning}.
Approaches bridging the indirect and direct paradigms are proposed in~\cite{iannelli2020structured,formentin2018core,dorfler2022bridging}.
Another successful approach in LQR, often deployed in an off-policy setting, is represented by policy-gradient methods, see the survey~\cite{hu2023toward}.
Complete characterizations of these methods for discrete-time LQR are in~\cite{bu2019lqr} and~\cite{fazel2018global}.
A model-free, gradient-based, strategy is proposed in~\cite{zhang2020policy}. 
In~\cite{mohammadi2021convergence}, the sample complexity and convergence properties for the continuous-time case are examined. 
In~\cite{mohammadi2020linear} the discrete-time case is considered. 
Sublinear regret bounds in model-free LQR are given in~\cite{abbasi2011regret, cohen2019learning}, while~\cite{cassel2020logarithmic,akbari2022achieving} provide poly-logarithmic regret bounds.
Sample complexity in model-free LQR is studied in~\cite{dean2020sample}.
Conversely, continuous-time on-policy techniques are proposed in~\cite{vrabie2009adaptive,jiang2012computational}.
In~\cite{possieri2022value}, stability guarantees on the learning dynamics are provided.
In the discrete-time context, the on-policy setting is addressed in~\cite{kiumarsi2015optimal} leveraging on policy iteration and value iteration approaches.
In~\cite{simchowitz2020naive}, regret bounds for online LQR are provided.
While the mentioned works about online approaches offer guarantees for (asymptotically and, some of them, probabilistically) obtaining stabilizing (possibly non-optimal) controllers, a thorough and explicit investigation into the stability properties of the closed-loop interlacing optimization, learning, and control tasks, governed by a time-varying and nonlinear dynamics, remains an open challenge.

Our main contribution is the development of a data-driven
on-policy control scheme with stability certificates in LQR for unknown systems.
Specifically, the estimated control policy is applied to the actual
(unknown) linear system, while it is concurrently refined toward the
optimal solution of the LQR problem.
The proposed method, termed \algname/, short for REcursive LEARNing policy gradient for LQR, relies on the so-called \emph{direct policy
  search} reformulation of the LQR problem, which is an optimization
problem with the control policy gain $\g$ being the decision
variable.
This optimization problem, with cost function parametrized by the
system matrices $(\As,\Bs)$, is addressed via a gradient-based method
combined with an estimation procedure to deal with the missing
knowledge of $(\As,\Bs)$.
In particular, the system matrices are progressively reconstructed via
a Recursive Least Squares (RLS) mechanism that iteratively processes
the state-input samples obtained from the actual, closed-loop system.
The on-policy nature of \algname/ stems from the fact that each
state-input sample is gathered by actuating the (yet non-optimal)
state feedback.
To ensure persistency of excitation, a probing dithering signal is
also fed into the (running) closed-loop dynamics.
The stability certificates for the learning and control closed-loop
system are proved by resorting to Lyapunov arguments and averaging
theory for two-time-scale systems.
Specifically, for the whole closed-loop system consisting of the
gradient update on the gain $\g$, the RLS scheme, and the system
dynamics, we show the exponential stability of a properly defined
steady state, in which: (i) the feedback policy is the optimal
solution of the LQR problem; (ii) the estimates of the unknown
matrices are exact; and (iii) the system state oscillates about the
origin with an amplitude arbitrarily tunable by setting the dither
magnitude.
These stability properties pave the way for characterizing algorithm effectiveness even in non-nominal conditions, where system and cost matrices change over time and/or disturbances affect the plant and the measurements.
In such scenarios, the closed-loop system would adapt dynamically, restoring optimality without requiring any restart of either the optimization or the learning process.
A key distinctive feature of our work is to provide a stability certificate for the overall closed-loop system that simultaneously addresses optimization, learning, and control tasks.
Most existing data-driven approaches (see, e.g.,~\cite{de2019formulas}) are not on-policy, namely, they are characterized by two distinct phases in which system samples are collected and then used to find the optimal gain.
Alternative, popular approaches (see, e.g.,~\cite{fazel2018global,krauth2019finite}) are based on improving the tentative policy by performing so-called experiments of the actuated system to evaluate it.
The main drawback of these approaches is that they are not online, namely, the real plant needs to be repeatedly initialized and actuated for a number of samples at each algorithm iteration.
Another branch of literature relies on the certainty equivalence principle, see, e.g.,~\cite{mania2019certainty,simchowitz2020naive}, which propose online strategies but focus on studying the incurred regret rather than the stability properties of the overall closed-loop system.

The paper unfolds as follows.
Section~\ref{sec:problem_setup_and_preliminaries} introduces the problem setup and some preliminaries.
Section~\ref{sec:algo} describes \algname/ and states its theoretical features.
Section~\ref{sec:analysis} analyzes the proposed scheme, while Section~\ref{sec:numerical_simulations} presents some numerical simulations.

\paragraph*{Notation}
The identity matrix in $\R^{n \times n}$ is $I_n$. 
The vector of zeros of dimension $n$ is denoted as $0_n$.
The vertical concatenation of $v_1, \dots, v_N$ is $\col(v_1, \dots, v_N)$.
$\cB_r(x) \!:=\! \{y \!\in\! \R^n \!\mid\! \norm{y \!-\! x} \!\leq\! r\}$ denotes the ball of radius $r \!>\! 0$ centered in $x \!\in\! \R^n$. %
$\sigma(A)$ denotes the spectrum of $A \in \R^{n \times n}$, while $A^\dagger$ denotes its Moore-Penrose inverse. %
We denote the Kronecker product by $\otimes$.
We denote the concatenation of the columns of $M \in \R^{n \times m}$ by $\vc{M} := \col([M]_{11},\dots,[M]_{n1},\dots,[M]_{1m},\dots,[M]_{nm}) \in \R^{nm}$, where $[M]_{ij}$ is the $(i,j)$-th entry of $M$.

\section{Preliminaries and Problem Setup}
\label{sec:problem_setup_and_preliminaries}

In this section we present some useful preliminaries and describe the problem we aim at investigating.

\subsection{Averaging theory for two-time-scale systems}
\label{app:averaging}

Consider the time-varying two-time-scale system
\begin{subequations}\label{eq:mixed_scale_system}
	\begin{align}
		\chitp & = \cA (\zt)\chit + h(\zt,\iter) + \epsilon g(\chit, \zt,\iter)
		\label{eq:mixed_scale_system_fast} 
		\\
		\ztp & = \zt + \epsilon f(\chit, \zt,\iter),
		\label{eq:mixed_scale_system_slow}
	\end{align}
\end{subequations}
with $\chit \in \R^{\nc}$, $\zt \in \R^{\nz}$, $g : \R^{\nc} \times \R^{\nz} \times \N \to \R^{\nc}$, $f: \R^{\nc} \times \R^{\nz} \times \N \to \R^{\nz}$, and $\cA: \R^{\nz} \to \R^{\nc \times \nc}$. 
Further, $\epsilon > 0$ is a tuning parameter that is useful to arbitrarily reduce the variations over time of subsystem~\eqref{eq:mixed_scale_system_slow}, which is therefore typically referred to as the \emph{slow} subsystem, with $\zt$ being the slow state.
Coherently, subsystem~\eqref{eq:mixed_scale_system_fast} is referred to as the \emph{fast} subsystem, with $\chit$ being the fast state. 
The analysis also relies on the investigation of a time-invariant auxiliary system associated to the slow dynamics, termed the \emph{averaged} system and defined as
\begin{align}\label{eq:average_app}
	\zatp = \zat + \epsilon f\av(\zat),
\end{align}
where $f\av:\R^\nz \to \R^\nz$ is given by
\begin{align}\label{eq:f_app}
	f\av(\z) := \lim_{T \to \infty}\tfrac{1}{T} \textstyle\sum_{\initer= \bar{t} +1}^{\bar{t} +T} f(0,\z,\initer).
\end{align}
The averaged system effectively ``neglects'' the time variability of $f(\cdot,\cdot,\iter)$ and assumes that the fast state $\chit$ is at the origin.
The underlying idea is that, as $\epsilon$ becomes smaller, a timescale separation can be established between the evolution of $\zt$ and the variations of $f(\cdot,\cdot,\iter)$ and $\chit$ so that the stability properties of~\eqref{eq:mixed_scale_system} can be inferred from those of~\eqref{eq:average_app}.
To this end, in the following, we report some conditions that will be used in the forthcoming algorithmic analysis.
First, we need the following regularity conditions on the vector fields of system~\eqref{eq:mixed_scale_system}.
\begin{assumption}\label{ass:lipschitz}
	There exists $r$ such that $f$, $g$, and $h$ are Lipschitz continuous over $\cB_{r}(0_{\nc + \nz})$.\oprocend
\end{assumption}
We also assume that the origin is an equilibrium of~\eqref{eq:mixed_scale_system}.
\begin{assumption}\label{ass:equilibrium}
	It holds $h(0,\iter) = 0$, $g(0,0,\iter) = 0$, and $f(0,0,\iter) = 0$ for all $\iter \in \N$.\oprocend
\end{assumption}
Third, we characterize the matrix function $\cA(\z)$ of~\eqref{eq:mixed_scale_system_fast}.
\begin{assumption}\label{ass:schur}
	There exist $r, m_1, m_2 \!>\! 0$ and $a_1, a_2 \in (0,1)$ such that, for all $\z \in \cB_{r}(0_{\nz})$ and $\iter \in \N$, it holds
	\begin{align*}
		m_1a_1^\iter \leq \snorm{ \cA(\z)^\iter} \leq m_2a_2^\iter.
	\end{align*}
	Further, $\cA$ is differentiable and it holds
	\begin{align*}
		\norm{\partial \cA(\z)/\partial \z_i} \leq k_a,
	\end{align*}
	for all $i \in \until{\nz}$, $\z \in \cB_{r}(0_{\nz})$, and some $k_a > 0$.\oprocend
\end{assumption}
Fourth, we impose that the vector field characterizing the averaged system, say $f\av:\R^\nz \to \R^\nz$, is well-posed.
\begin{assumption}\label{ass:limit}
The function $f$ is piecewise continuous in $\iter$ and the limit $\lim_{T \to \infty}\tfrac{1}{T} \textstyle\sum_{\initer= \bar{t} +1}^{\bar{t} +T} f(0,\z,\initer)$
exists uniformly in $\bar{\iter} \in \N$ and for all $\z \in \cB_r(0_{\nz})$.\oprocend
\end{assumption}
Finally, we characterize the difference between $f$ and $f\av$. %
\begin{assumption}\label{ass:nu}
	Consider $f\av$ as defined in~\eqref{eq:f_app} and let $\Delta f: \R^{\nz} \times \N \to \R^{\nz}$ be defined as
	\begin{align*}
		\Delta f(\z,\iter) := f(0,\z,\iter) - f\av(\z).
	\end{align*}
	Then, there exists a nonnegative strictly decreasing function $\nu(\iter)$ such that $\lim_{\iter \to \infty} \nu(\iter) = 0$ and
	\begin{subequations}%
		\begin{align*}
			\norm{\tfrac{1}{T}\textstyle\sum_{\initer = \bar{\iter} + 1}^{\bar{\iter} + T}\Delta f(\z,\initer)} &\leq \nu(T)\norm{\z}
			\\
			\norm{\tfrac{1}{T}\textstyle\sum_{\initer = \bar{\iter} + 1}^{\bar{\iter} + T}\frac{\partial\Delta f(\z,\initer)}{\partial \z}} &\leq \nu(T),
		\end{align*}
	\end{subequations}
	uniformly in $\bar{\iter} \in \N$ and for all $\z \in \cB_{r}(0_{\nz})$.
	\oprocend
\end{assumption}
We are ready to provide a stability result for~\eqref{eq:mixed_scale_system}. 
Essentially, for sufficiently small $\epsilon$, exponential stability of the origin for~\eqref{eq:average_app} is enough to get exponential stability of the origin for the interconnected time-varying system~\eqref{eq:mixed_scale_system}.
\begin{theorem}\label{th:bai}\cite[Theorem~2.2.4]{bai1988averaging}
	Consider system~\eqref{eq:mixed_scale_system} and let Assumptions~\ref{ass:lipschitz},~\ref{ass:equilibrium},~\ref{ass:schur},~\ref{ass:limit} and~\ref{ass:nu} hold.
	If there exists $\epsilon\av$ such that, for all $\epsilon \in (0,\epsilon\av)$, the origin is exponentially stable for system~\eqref{eq:average_app}, then there exists $\bar{\epsilon} \in (0,\epsilon\av)$ such that, for all $\epsilon \in (0,\bar{\epsilon})$, the origin is an exponentially stable equilibrium of system~\eqref{eq:mixed_scale_system}.
	\oprocend
\end{theorem}
	
\subsection{On-Policy Data-Driven LQR: Problem Setup}
\label{sec:problem_statement}

In this paper, we focus on the LQR problem
\begin{subequations}	
	\label{eq:problem}
	\begin{align}
		\min_{\substack{\x_1, \x_2,\ldots,\\ \uu_0,\uu_1,\ldots}} \: \: & \: \bE\bigg[\halff\sum_{\timeid=0}^{\infty} \Big( {\xtime}\T Q \xt + {\utime}\T R \ut  \Big)\bigg]
		\\
		\subj \: & \: \xtimep = \As \xtime + \Bs \utime, \qquad \qquad \x\ud0 \sim \cX\ud0,
		\label{eq:dyn}
	\end{align}
\end{subequations}	
where  $\xtime \in \R^\dimx$ and $\utime \in \R^\dimu$ denote, respectively, the state and the input of the system at time $\timeid \in \N$, while $\As \in \R^{\dimx \times \dimx}$ and $\Bs \in \R^{\dimx \times \dimu}$ are the state and input matrices.
As for the initial condition $\x\ud0 \in \R^\dimx$, we assume that it is drawn from a (known) probability distribution $\cX\ud0$.
Hence, the operator $\bE[\cdot]$ denotes the expected value with respect to $\cX\ud0$.
The cost matrices $Q \in \R^{\dimx \times \dimx}$ and $R \in \R^{\dimu \times \dimu}$ satisfy $Q = Q\T > 0$ and $R = R\T > 0$.
In case of detectability of $(\As,Q_0)$ with $Q:=Q_0\T Q_0$, the positive definiteness of $Q$ can be relaxed to mere positive semidefiniteness.
We characterize $(\As,\Bs)$ as follows.
\begin{assumption}[Unknown System Properties]
	\label{ass:unknown}
	The pair $(\As,\Bs)$ is unknown and controllable. 
	\oprocend
\end{assumption}
As it will be useful later, we collect the pair $(\As,\Bs)$ in a single variable $\pr :=
	[\As\; \Bs]\T \in \dpr$.
	It is well-known that the optimal solution to problem~\eqref{eq:problem} is given by
a linear time-invariant policy $\utime = \gstar \xtime$ with the optimal gain $\gstar \in \R^{\dimu \times \dimx}$ given by
	\begin{align}\label{eq:DARE}
		\gstar = - (R + \Bs\T P_\star\Bs)\inv \Bs\T P_\star\As,
	\end{align}
where $P_\star\in \R^{\dimx \times \dimx}$ solves the Discrete-time Algebraic Riccati Equation associated to problem~\eqref{eq:problem}, see~\cite{anderson2007optimal}.%

In this work, we are interested in devising a data-driven on-policy strategy to get a state-feedback controller solution to~\eqref{eq:problem}.
Thus, the problem can be posed as 
designing a learning and control scheme that is capable of 
\begin{enumerate}[label=(\roman*)]
	\item learning the optimal policy solution to problem~\eqref{eq:problem},
	\item estimating the unknown system matrices,
	\item actuating the (real) system with the currently available state-feedback policy,
\end{enumerate}
while ensuring asymptotic stability properties on the closed-loop learning and control system.

\begin{figure}[htpb]
	\centering
	\includegraphics[scale=\scalebd]{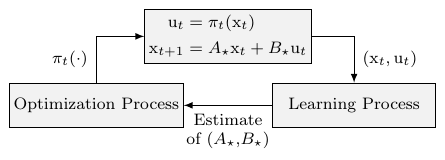}
	\caption{Schematic representation of the stability-certified on-policy LQR setup.}
	\label{fig:scheme_intro}
\end{figure}

As shown in Fig.~\ref{fig:scheme_intro}, our scheme iteratively computes a tentative policy, say $\pi\ud\iter(\cdot)$, to be actuated on the real, unknown system. 
One single sample of the system state is then collected and fed into a learning mechanism refining the matrices estimates, which, in turn, are used to improve the policy.
The distinctive feature of our approach is that these steps are interwoven rather than temporally separate, as is typically done in the literature.

\subsection{Model-based Gradient Method for LQR}

Next, we recall the key ingredients for devising a model-based gradient method to address problem~\eqref{eq:problem}.

\subsubsection{Model-based reduced problem formulation}

First of all,  we recall an equivalent (unconstrained) formulation of problem~\eqref{eq:problem} that explicitly imposes the linear feedback structure to the optimal input and is amenable for gradient-based algorithmic solutions.
Problem~\eqref{eq:problem} is rewritten by substituting in the dynamics and in the cost function the input in linear feedback form $\utime = \g \xtime$,
where $\g \in \R^{\dimu \times \dimx}$ is to be computed.
Hence, for all $\timeid \in \N$, the state is uniquely determined as 
\begin{align}
	\label{eq:state_at_t}
	\xtime = (\As + \Bs\g)^\timeid \x\ud0, \qquad \x\ud0 \sim \cX\ud0.
\end{align}
By using~\eqref{eq:state_at_t}, assuming, without loss of generality, that $\cX\ud0$ is a uniform distribution on the unit sphere, and taking the expectation on $\x\ud0$, we rewrite problem~\eqref{eq:problem} as %
\begin{align}\label{eq:LQR}
	\min_{\g \in \cK}%
	\: & \: \J(\g,\pr),
\end{align} 
where $\cK \! := \! \{\! \g \!\in\! \R^{m \times n} \!\!\mid\!\! \As \! + \! \Bs\g \text{ is Schur}\} \!\subseteq\! \R^{m \times n}$ is the stabilizing gains' set and $\J\!\!: \!\cK \!\times\! \dpr \!\to\! \R$ reads as 
\begin{align}
	& \J(\g,\pr)=
	\halff
	\Tr\left( 
	\sum_{\timeid=0}^\infty
	(\As + \Bs\g)^{\timeid,\top}
	(Q+\g\T R \g)
	(\As + \Bs\g)^{\timeid}\right).\label{eq:J}
\end{align}
This formulation highlights that (i) the overall problem actually depends on the gain $\g$ only, and, (ii) the optimal gain $\gstar$ does not depend on the initial condition $\x\ud0$.
\begin{remark}
	Due to the linearity of the expected value operator and the properties of the trace, the formulation in~\eqref{eq:J} holds up to a constant scaling factor for any probability distribution $\mathcal{X}_0$ with a well-defined second moment.
\end{remark}

\subsubsection{Model-based gradient method for problem~\eqref{eq:LQR}}

The set of stabilizing gains $\cK$ is open~\cite[Lemma~IV.3]{bu2020topological} and connected~\cite[Lemma~IV.6]{bu2020topological}.
Moreover, the cost function $\J(\cdot,\pr)$ is coercive~\cite[Lemma~3.7]{bu2019lqr}.
Had the pair $(\As,\Bs)$ been known, the gradient descent method could have been used to solve problem~\eqref{eq:LQR} (see, e.g.,~\cite{bu2019lqr}).
Namely, at each iteration $\iter \in \N$, an estimate $\gt$ of $\gstar$ is maintained and iteratively updated according to
\begin{align}
	\gtp = \gt - \ssz \G(\gt,\pr),
	\label{eq:desired_update}
\end{align}
where $\ssz > 0$ is the stepsize and $\G: \R^{m \times n} \times \dpr \to \R^{m \times n}$ is the gradient of $J$ with respect to $\g$ evaluated at $(\gt,\pr)$, when $\R^{m \times n}$ is equipped with the Frobenius inner product.
By initializing $\g\ud0 \in \cK$ and selecting a proper stepsize $\ssz$, the optimal gain $\gstar$ is an exponentially stable equilibrium of system~\eqref{eq:desired_update}, see~\cite[Theorem~4.6]{bu2019lqr}.
Given $\g \in \cK$, we note that $\G(\g,\pr)$ reads as
\begin{align}
	\label{eq:explicit_form_gradient}
	\!\!\!\!\!&\G(\g,\pr) \! = \! \left( R \g +\Bs\T P (\As + \Bs\g) \right) W^c,
\end{align}
where $W^c, P \in \R^{n \times n}$ are the solutions to the equations
\begin{align*}%
	&
	(\As + \Bs\g) W^c (\As + \Bs\g)\T - W^c=-I_n
	\\
	&
	(\As + \Bs\g)\T P (\As + \Bs\g) - P = -(Q+{\g}^\top R \g).
\end{align*}
However, since our goal is to address the problem setup described in Section~\ref{sec:problem_statement}, the pair $(\As,\Bs)$ is unknown.
Consequently, the update~\eqref{eq:desired_update} cannot be implemented.

\section{On-policy LQR for Unknown Systems: Concurrent Learning and Optimization}
\label{sec:algo}

\begin{figure}[H]
	\centering
	\includegraphics[scale=\scalebd]{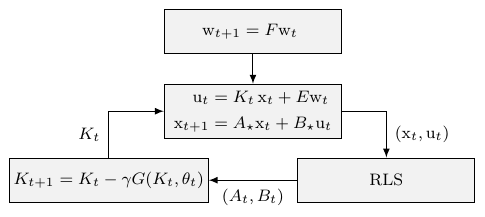}
	\caption{Representation of the concurrent learning and optimization scheme implemented by \algname/.}
	\label{fig:scheme}
\end{figure}

In this section, we present \algname/, a concurrent learning and optimization algorithm developed to solve the stability-certified on-policy LQR setup described in Section~\ref{sec:problem_statement}.
The proposed on-policy strategy feeds the real system at each iteration
$\iter$ with the current feedback strategy including also an exogenous dithering signal $\wt$.
Then, a new data sample from the system is collected and used to improve the estimates $(\At,\Bt)$ of the unknown $(\As,\Bs)$ via a learning process
inspired by Recursive Least Squares (RLS).
In turn, $(\At,\Bt)$ is used to refine the feedback gain $\gt$ according to the (approximated) gradient method. %
Fig.~\ref{fig:scheme} shows the overall scheme.
Our \algname/ strategy is reported in Algorithm~\ref{alg:algorithm}, where $\est \in \dpr$ denotes the estimate of $\pr$ at iteration $\iter$
and $(\At,\Bt) \in \R^{n\times n} \times \R^{n \times m}$ are the corresponding estimates of $(\As, \Bs)$. %
Further, $\hst \in \R^{(n+m)\times(n + m)}$ and $\rrt \in \R^{(n+m)\times n}$ are two additional states of the learning part and $\lambda \in (0,1)$ is a forgetting factor. %
\begin{algorithm}[H]
	\begin{algorithmic}%
		\For{$\iter = 0, 1, 2 \ldots$}

		\State \textbf{Data collection:}
		generate
		\vspace{-2ex}
		\begin{align*}
				\wtp &= \Sw\wt
				\\
				\drt & = \EE\wt
		\end{align*}
		\State and actuate
		\vspace{-2ex}
		\begin{align*}
				\ut &= \gt \xt + \drt
				\\
				\xtp &=  \As \xt + \Bs \ut
				\\
				\mst & = \xtp\T
		\end{align*}
		
		\State \textbf{Learning process:}
		compute
		\begin{subequations}\label{eq:learning_process}
			\begin{align}
				\hstp & = \lambda\hst + 
				\begin{bmatrix}
					\xt \\ \ut
				\end{bmatrix}
				\begin{bmatrix}
					\xt \\ \ut
				\end{bmatrix}\T
				\label{eq:hs_update}
				\\
				\rrtp & = \lambda\rrt + \begin{bmatrix}
					\xt\\
					 \ut
				\end{bmatrix} \mst\label{eq:rr_update}
				\\
				\estp &= \est - \ssz \hst^\dagger\left(\hst\est - \rrt\right)
				\label{eq:es_update}
			\end{align}
		\end{subequations}
		
		\State \textbf{Optimization process:}
		update
		\begin{align}
			\gtp = \gt - \ssz \G(\gt,\est)\label{eq:g_update}
		\end{align}
		\EndFor
		\caption{\algname/}
		\label{alg:algorithm}
	\end{algorithmic}
\end{algorithm}
Next, we detail the main steps of the proposed algorithm.	

\paragraph*{Data collection}
	Data from the controlled system~\eqref{eq:dyn} are recast in an identification-oriented form described by
	\begin{align} \label{eq:true_model_open_loop}
		\underbrace{\xtp\T}_{\displaystyle \mst} 
		= 
		\underbrace{
			\begin{bmatrix}
				\xt\T& \ut\T%
		\end{bmatrix}}_{\displaystyle \rg(\xt,\ut)\T%
	}
		\underbrace{
			\begin{bmatrix}\As &\Bs \end{bmatrix}\T}_{\displaystyle \pr}.
	\end{align}

\paragraph*{Learning process}
	
The adopted learning strategy to compute an estimate of $\pr$ relies on the interpretation of the least squares problem as an online optimization.
Specifically, with the measurements~\eqref{eq:true_model_open_loop} at hand, we consider, at each $\iter \in \N$, the online optimization problem
\begin{align} \label{eq:online_problem}
	\min_{\es \in \dpr} \halff\textstyle\sum_{\initer=0}^{\iter}\lambda^{\iter - \initer}\norm{\rg(\x\ud \initer,\uu\ud \initer)\T\es - \ms\ud \initer}^2.
\end{align}
We iteratively address~\eqref{eq:online_problem} by updating a solution estimate $\est \in \dpr$ through a ``scaled'' gradient method with Newton-like scaling matrix, namely
\begin{align*} 
	\estp &= \est
	- \ssz (\textstyle\sum_{\initer=0}^{\iter}\lambda^{\iter - \initer}\chs(\x\ud \initer,\uu\ud\initer))^{\dagger}
	\textstyle\sum_{\initer=0}^{\iter} 
	\lambda^{\iter - \initer}(\chs(\x\ud\initer,\uu\ud\initer)\est - \crr(\x\ud\initer,\uu\ud\initer,\ms\ud\initer))),
\end{align*}
where $\chs(\x\ud\initer,\uu\ud \initer)$ and $\crr(\x\ud\initer,\uu\ud \initer,\ms\ud\initer)$
are defined as
\begin{align*}
	\chs(\x\ud\initer,\uu\ud \initer) &:= \rg(\x\ud\initer,\uu\ud \initer)\rg(\x\ud\initer,\uu\ud \initer)\T
	\\
	\crr (\x\ud\initer,\uu\ud \initer,\ms\ud\initer) &:= \rg(\x\ud\initer,\uu\ud\initer)\ms\ud\initer.
\end{align*}
To avoid storing the whole history of $\chs$ and $\crr$, 
we iteratively track them through the matrix states $\hst \in \R^{(n + m) \times (n + m)}$ and $\rrt \in \R^{(n + m) \times n}$ 
giving rise to~\eqref{eq:learning_process}.

\paragraph*{Optimization process}

The estimate $\est$ is concurrently used in the update of the gain $\gt$, replacing the unavailable $\pr$ into~\eqref{eq:desired_update} giving rise to~\eqref{eq:g_update}.
To ensure sufficiently informative data, we equip our feedback policy with an additive dithering signal $\drt \in \R^{m}$, namely
\begin{align}
	\ut = \gt\xt + \drt,\label{eq:control_law}
\end{align}
where $\drt \in \R^{m}$ is the output of an exogenous, 
discrete-time oscillator dynamics (see, e.g.,~\cite{turner2003recursive}) described by
\begin{subequations}\label{eq:drt_dynamics}
\begin{align}
	\wtp &= \Sw\wt
	\\
	\drt &= \EE\wt,
\end{align}
\end{subequations}
where $\wt \in \R^{\nw}$, with $\nw \ge \dimx +\dimu$, is the state, while $\Sw \!\in \! \R^{\nw \times \nw}$ and $\EE \! \in \! \R^{\dimu \times \nw}$ are state and output matrices. 
The design requirements of~\eqref{eq:drt_dynamics} are as follows.
\begin{assumption}[Persistency of Excitation]\label{ass:dither}
	The signal $\wt$ is \emph{persistently exciting}, while $\drt$ is \emph{sufficiently rich of order $(\dimx + 1)$}, i.e., there exist $\alpha_1, \alpha_2, \per, \tPE > 0$ such that, if $\w\ud 0\neq 0_{\nw}$, then, for all $\bar{\iter} \in \N$, it holds
	\begin{subequations}\label{eq:PE_all}%
	\begin{align}%
&		\alpha_1 I_{\nw} 
		\le 
		\textstyle\sum_{\initer=\bar{\iter}+1}^{\bar{\iter} + \per}\w\ud \initer \w\ud\initer\T
		\le
		\alpha_2 I_{\nw}, 
		\label{eq:PE}
		\\
		\label{eq:PE_order}
&		\rank\left(\begin{bmatrix}
			\dr\ud{\bar{\iter}}& \dr\ud{\bar{\iter}+1}& \hdots& \dr\ud{\bar{\iter}+\tPE - n - 1} 
			\\
			\vdots& \vdots& \ddots& \vdots 
			\\
			\dr\ud{\bar{\iter}+n}& \dr\ud{\bar{\iter}+n+1}& \hdots& \dr\ud{\bar{\iter}+\tPE - 1} 
		\end{bmatrix}\right) 
		= m (n+1),
	\end{align}
\end{subequations}%
	Further, the eigenvalues of $\Sw$ lie on the unit circle.
	\oprocend
\end{assumption}

In finite-time windows, the property~\eqref{eq:PE_order} is known as \emph{persistency of excitation of order $(n+1)$} (see~\cite{willems2005note,de2019formulas}), while we used the notion of \emph{sufficient richness}~\cite{bai1985persistency}.
\begin{remark}\label{rem:F_E}
	A possible way to build $F$ and $E$ to verify Assumption~\ref{ass:dither} is as follows.
First, we set $\harm := n +1$, $\nw = 2\harm$, and $\Sw := \blkdiag(\Sw_1,\dots,\Sw_{\harm}),$
where, for all $i \in \until{\harm}$, $\Sw_{i} \in \R^{2\times 2}$ is defined as
\begin{align*}
	\Sw_{i} := 
	\begin{bmatrix}
		\cos(\omega_i)& \sin(\omega_i)
		\\
		-\sin(\omega_i)& \cos(\omega_i)
	\end{bmatrix},
\end{align*}
for given $\omega_i$ such that $\omega_i = 2\omega_{i-1}$ for all $i \in \{2,\ldots,\harm\}$.
By choosing an initial condition $\w\ud 0$ that satisfies
\begin{align*}%
	([\w\ud{0}]_{2i -1})^2 + ([\w\ud{0}]_{2i})^2 \ne 0, \quad i \in \until{\harm},
\end{align*}
the chosen structure of $\Sw$ guarantees~\eqref{eq:PE} according to~\cite[Thm.~2]{padoan2017geometric}.
As for~\eqref{eq:PE_order}, it is achieved by selecting $E$ such that $[
E\T\; (E\Sw)\T\; \hdots\; (E\Sw^{n})\T]$ is nonsingular.\oprocend
\end{remark}
The closed-loop system resulting from Algorithm~\ref{alg:algorithm} is
\begin{subequations}%
\label{eq:algo_closed_loop}%
\begin{align}%
	\wtp &= \Sw\wt\label{eq:algo_closed_loop_drt}
	\\
	\xtp &= (\As + \Bs\gt)\xt +  \Bs\EE\wt\label{eq:algo_closed_loop_xt}
	\\
	\hstp & = \lambda\hst +
	\begin{bmatrix}
		\xt \\
		\gt\xt + \EE\wt
	\end{bmatrix}
	\begin{bmatrix}
		\xt \\
		\gt\xt \! + \! \EE\wt
	\end{bmatrix}	\T
	\label{eq:algo_closed_loop_hst}
	\\
	\rrtp & = \lambda\rrt +
	\begin{bmatrix}
		\xt \\
		\gt\xt \! + \! \EE\wt
	\end{bmatrix}
	\begin{bmatrix}
		\xt \\
		\gt\xt \! + \! \EE\wt
	\end{bmatrix} \T
	\pr
	\label{eq:algo_closed_loop_rrt}
	\\
	\estp &= \est - \ssz\hst^\dagger\left(\hst\est - \rrt\right)
	\label{eq:algo_closed_loop_est}
	\\
	\gtp &= \gt - \ssz\G(\gt,\est),\label{eq:algo_closed_loop_gt}
\end{align}
\end{subequations}
in which we use the expressions of $\mst$ (cf.~\eqref{eq:true_model_open_loop}) and $\ut$ (cf.~\eqref{eq:control_law}).
In order to establish the stability properties of the closed-loop system~\eqref{eq:algo_closed_loop}, let us introduce the sets $\cS := R^{\nw} \times \R^{n} \times \R^{(n + m)\times(n+m)} \times \R^{(n + m) \times n} \times \dpr \times \cK$ and
	$\Vs(\Pi_1,\Pi_2,\Pi_3)
	   := \{(w,x,\hs,\rr,\es,\g) \in \cS
		\mid  \w \ne 0_{\nw},
		x = \Pi_1\w, 
		\hs = \opH(\Pi_2, \w), 
		\rr = \opS(\Pi_3, \w), 
		(\es,\g) = (\pr,\gstar)\}$,
	where $\opH: \R^{(n+m)^2 \times \nw^2} \times \R^{\nw}\to \R^{(n+m) \times (n+m)}$ and $\opS: \R^{(n + m)m \times \nw^2} \times \R^{\nw}\to\R^{(n + m) \times n}$ are defined as
\begin{subequations}\label{eq:oph_ops}
	\begin{align}
		\opH(\Pi_1, \w) &:= \unvc{\Pi_1\vc{\w\w\T}}
		\label{eq:v_h}
		\\
		\opS(\Pi_2,\w) &:= \unvc{\Pi_2\vc{\w\w\T}}.
		\label{eq:v_s}
	\end{align}
\end{subequations}
\begin{theorem}\label{th:convergence}
	Let Assumptions~\ref{ass:unknown} and~\ref{ass:dither} hold.
	Then, %
	there exist $\pix \in \R^{n\times \nw}$, $\pih \in \R^{(n+m)^2 \times \nw^2}$, $\pir \in \R^{(n + m)m \times \nw^2}$, and $\bar{\ssz} > 0$ such that, for all $\ssz \in (0,\bar{\ssz})$, the set $\Vs(\pix,\pih,\pir)$ is exponentially stable for system~\eqref{eq:algo_closed_loop}. %
	\oprocend
\end{theorem}
The proof of Theorem~\ref{th:convergence} is provided in Section~\ref{sec:proof_of_convergence}.

Besides stability, %
Theorem~\ref{th:convergence} ensures exponential convergence of $(\xt,\est,\gt)$ toward $(\pix\wt,\pr,\gstar)$.
Namely, for some $a_1,a_2 \!> 0$, along the trajectories of~\eqref{eq:algo_closed_loop}, it holds
\begin{align}
		&\norm{\col(\xt  \! - \!  \pix\wt,\est \! - \! \pr,\gt \! - \! \gstar)} 
		 \leq a_1\exp(-a_2\iter),\!\label{eq:exp} 
	\end{align}
for all $\iter\in\N$.
Hence, our method asymptotically reconstructs $(\As,\Bs)$ and $\gstar$ with linear rate.
Moreover, we recall that $\wt$ follows an oscillating dynamics (cf. Assumption~\ref{ass:dither}) such that $\norm{\wt} = \norm{\w\ud{0}}$ for all $\iter \in \N$.
Then, by~\eqref{eq:exp} and for all $\rho > \|\pix\w\ud0\|$, the ball $\cB_{\rho}(0_\dimx)$ is exponentially attractive for~\eqref{eq:algo_closed_loop_xt}. %
Since we can arbitrarily reduce $\w\ud0$, this implies that the origin of~\eqref{eq:algo_closed_loop_xt} is practically exponentially stable, provided that the other states lie in their steady-state locus.
\begin{remark}
	From the proof of Theorem~\ref{th:convergence} (which uses Proposition~\ref{prop:average} proved in Appendix~\ref{app:proof}), one can see that the initial condition $(\es\ud0,\g\ud0)$ must lie in a neighborhood of $(\pr, \gstar)$.
	Hence, the initialization requirements for \algname/ in Theorem~\ref{th:convergence} are more stringent than those in existing results in the literature, which typically only require an initial stabilizing controller for the true model $(\As,\Bs)$ (see~\cite{qin2014online,modares2016optimal,krauth2019finite,pang2021robust,lopez2023efficient,fazel2018global,zhang2020policy,hu2023toward}).
	These works, however, only prove convergence to a neighborhood of $\g^\star$ and do not analyze the stability of the closed-loop system arising from the interaction between the real system and the proposed algorithm.
	By contrast, Theorem~\ref{th:convergence} guarantees \emph{exact} convergence to the optimal gain $\g^\star$, in addition to stability of the closed-loop system~\eqref{eq:algo_closed_loop}. 
	\oprocend
\end{remark}
\begin{remark}
	In realistic scenarios, an approximate (controllable) model of the system is typically available and can be used as $(A\ud{0},B\ud{0})$.
	One can then compute a stabilizing controller $\g\ud{0}$ for the pair $(A\ud{0},B\ud{0})$.
	For instance, $\g\ud{0}$ can be obtained as the solution of the discrete-time algebraic Riccati equation~\eqref{eq:DARE} with $(A\ud{0},B\ud{0})$ in place of the unknown $(A_\star,B_\star)$.
	\oprocend
\end{remark}
\begin{remark}
	The proof of Theorem~\ref{th:convergence} exploits system theory tools based on averaging theory for two-time-scale systems (cf.~Theorem~\ref{th:bai} in Section~\ref{app:averaging}) and, thus, introduces an auxiliary system called the \emph{averaged system}. 
	Such auxiliary system involves modified averaged dynamics of $\est$ and $\gt$ (see Section~\ref{sec:average_system}) and is shown to have an exponentially stable equilibrium in its origin.
	Further, we are also able to show that $\gt$ remains a stabilizing gain for $(\As,\Bs)$ at all $\iter \! \in \! \N$ (see Appendix~\ref{app:proof}).
	Such stabilizing property combined with closeness between trajectories of the averaged and original systems (imposed through $\ssz$, see, e.g.,~\cite{bai1988averaging}), allows for concluding that $\gt$, generated by~\eqref{eq:algo_closed_loop}, stabilizes $(\As,\Bs)$ for all $\iter \in \N$.
	 \oprocend
\end{remark}

\section{Stability Analysis}
\label{sec:analysis}
	
In this section, we perform the stability analysis of the closed-loop system~\eqref{eq:algo_closed_loop}. %
First, we rewrite it in suitable error coordinates. 
Second, we resort to the averaging theory to prove the exponential stability of the origin for the averaged system associated to the error dynamics.
This result is then exploited to prove Theorem~\ref{th:convergence}.
\subsection{Closed-Loop Dynamics in Error Coordinates}

As a preliminary step, we express system~\eqref{eq:algo_closed_loop} into suitable error coordinates.
First, we consider vectorized versions of the matrix updates in~\eqref{eq:algo_closed_loop_hst}-\eqref{eq:algo_closed_loop_rrt}. 
To this end, let $\vhs \in \R^{(n+m)^2}$ and $\vrr \in \R^{(n+m)n}$ be defined as
\begin{align} \label{eq:vectorized}
	\begin{bmatrix}
		\hs
		\\
		\rr
	\end{bmatrix}
	 \longmapsto
	\begin{bmatrix}
		\vhs
		\\
		\vrr
	\end{bmatrix} := \begin{bmatrix}
		\vc{\hs}
		\\
		 \vc{\rr}
	\end{bmatrix}.
\end{align}
Therefore, \eqref{eq:algo_closed_loop_hst}-\eqref{eq:algo_closed_loop_rrt} can be recast as
\begin{subequations}
		\label{eq:algo_closed_loop_with_vc}
		\begin{align}
			\vhstp &=  \lambda\vhst  + 
			\ssvc{
				\begin{bmatrix}
					\xt \\ \gt\xt  +  \EE\wt
				\end{bmatrix}
				\begin{bmatrix}
					\xt \\ \gt\xt  +  \EE\wt
				\end{bmatrix}^{\top}
			}
			\label{eq:algo_closed_loop_with_vc_hs}
			\\
			\vrrtp  &=  \lambda\vrrt  +  
			\ssvc{
				\begin{bmatrix}
					\xt \\\gt\xt  +  \EE\wt
				\end{bmatrix}
				\begin{bmatrix}
					\xt \\ \gt\xt  +  \EE\wt
				\end{bmatrix}^{\top}
				\pr
			}.
		\end{align}
\end{subequations}

Next, we will inspect~\eqref{eq:algo_closed_loop_with_vc} together with~\eqref{eq:algo_closed_loop_xt} to provide the steady-state locus (see, e.g.,~\cite[Ch.~12]{isidori2017lectures} for a formal definition) when the system is fed with the signal $\wt$, which evolves according to~\eqref{eq:algo_closed_loop_drt}.
To this end, set $\nc := n + (n+m)^2 + (n+m)n$ and let $\chi \in \R^{\nc}$ be defined as 
\begin{align*}
	\chi := \col(\x,\vhs,\vrr).
\end{align*}
Then, by using~\eqref{eq:algo_closed_loop_with_vc}, the dynamics in~\eqref{eq:algo_closed_loop_drt}-\eqref{eq:algo_closed_loop_rrt} can be compactly expressed in the new coordinates as
\begin{subequations}
	\label{eq:compact_system}
	\begin{align}
		\wtp &=\Sw\wt
		\\
		\chitp &= \cAc(\gt)\chit + \fun(\chit,\gt,\wt),
	\end{align} 
\end{subequations}
where we introduced $\cAc: \R^{m \times n} \to \R^{\nc \times \nc}$ and $\fun: \R^{\nc} \times \R^{m \times n} \times \R^{\nw} \to \R^{\nc}$ be defined as 
\begin{subequations}\label{eq:A_k_phi}
\begin{align}
	\cAc(\g) &:= 
	\begin{bmatrix}
		\As + \Bs\g& 0&
		\\
		0& \lambda I_{(\dimx+\dimu)(2\dimx+\dimu)}
	\end{bmatrix}\label{eq:cAc}
	\\
	\fun(\chi,\g,\w) &:= \begin{bmatrix}
		\Bs\EE\w 
		\\
		\vc{
			\begin{bmatrix}
				\chi_1
				\\
				\g\chi_1 \! + \! \EE\w
			\end{bmatrix}
			\begin{bmatrix}
				\chi_1 
				\\ 
				\g\chi_1 \! + \! \EE\w
			\end{bmatrix} \T
		}
		\\
		\vc{
			\begin{bmatrix}
				\chi_1 
				\\
				\g\chi_1 \! + \! \EE\w
			\end{bmatrix}
			\begin{bmatrix}
				\chi_1 
				\\ 
				\g\chi_1 \! + \! \EE\w
			\end{bmatrix} \T
			\pr
		}
	\end{bmatrix},\label{eq:phi}
\end{align}
\end{subequations}
in which $\chi_1 \in \R^{\dimx}$ denotes the first $\dimx$ components of $\chi$.

System~\eqref{eq:compact_system} together with the exosystem~\eqref{eq:algo_closed_loop_drt} is a cascade whose steady-state locus can be characterized by the nonlinear map $\sts: \R^{\nw} \to \R^{\nc}$ defined as 
\begin{align}\label{eq:ss}
	\sts(\w) := \begin{bmatrix}
		\pix\w
		\\
		\begin{bmatrix}
			\pih& 0
			\\
			0&\pir
		\end{bmatrix}\vc{\w\w\T}
	\end{bmatrix},
\end{align}
where $\pix$, $\pih$, and $\pir$ are those referred in Theorem~\ref{th:convergence} (see~\eqref{eq:pix_equation} and~\eqref{eq:sylvester_vec} in Appendix~\ref{app:ss} for their explicit definition).
Formally, the following lemma holds true.
\begin{lemma}\label{lemma:ss}
	Let the assumptions of Theorem~\ref{th:convergence} hold true.
  Consider the map $\sts$ defined in~\eqref{eq:ss}, the feedback gain $\gstar$ solving~\eqref{eq:problem}, the matrix $\F$ as in~\eqref{eq:drt_dynamics}, and the functions $\cAc$ and $\fun$ defined in~\eqref{eq:A_k_phi}.
	Then, it holds 
	\begin{align}
		\sts(\Sw\w) = \cAc(\gstar)\sts(\w) + \fun(\sts(\w),\gstar,\w),\label{eq:steady_state_locus}
	\end{align}
	for all $\w \in \R^{\nw}$.
	Moreover, it holds
	\begin{align}
		({\pr}\T \otimes I_{n + m})\pih 
		 = \pir.%
		\label{eq:pih_pir_pr}
	\end{align}
	\oprocend
\end{lemma}
The proof of Lemma~\ref{lemma:ss} is provided in Appendix~\ref{app:ss}.

Lemma~\ref{lemma:ss} ensures that $\col(\sts(\w),\pr,\gstar)$ is the steady-state locus of the overall closed-loop system~\eqref{eq:algo_closed_loop}. 
In this regard, we also include condition~\eqref{eq:pih_pir_pr} since it allows us to show that $\pr$ is an equilibrium of~\eqref{eq:algo_closed_loop_est} restricted to the case in which $\hst$ and $\rrt$ lie in the steady-state locus.
Indeed, when $\chit = \sts(\wt)$, system~\eqref{eq:algo_closed_loop_est} reduces to
\begin{align*}
	&\estp\Big|_{\chit = \sts(\wt)} 
	= 
	\est - \ssz\left(\hst\est - \rrt\right)\Big|_{\chit = \sts(\wt)}
	\notag
	\\
	&
	=
	\est - \ssz\opS(\pir,\wt)\est
	\notag\\
	&
	=
	\est - \ssz\unvc{\left({\pr}\T \otimes I_{n + m}\right)\pih\vc{\wt\wt\T}}
	\\
	&
	= \est - \ssz\opH(\pih, \wt)\left(\est - \pr\right),
\end{align*}
where we use a vectorization operator property\footnote{Given any two matrices $X_1 \in \R^{n_1 \times n_2}$ and $X_2 \in \R^{n_2 \times n_3}$, it holds $\vc{X_1X_2} = (X_2\T \otimes I_{n_1})\vc{X_1}$.} and the definitions of $\opH$ and $\opS$ given in~\eqref{eq:oph_ops}.
As for the equilibrium of~\eqref{eq:algo_closed_loop_gt} when the other states lie on the steady-state locus, it turns out to be $\gstar$ since $\G(\gstar,\pr) = 0$.

\begin{figure}[!htpb]
	\centering
	\includegraphics[scale=\scalebd]{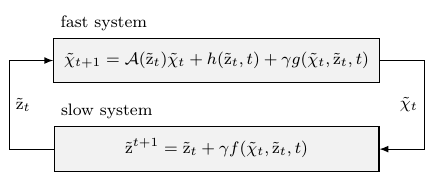}
	\caption{Block diagram describing system~\eqref{eq:mixed_scale_system_our}.}
	\label{fig:block_diagram}
\end{figure}

Before proceeding, let us collect also the remaining states in~\eqref{eq:algo_closed_loop} in $\z \in \R^{\nz}$, with $\nz := (n+2m)\times n$, defined as 
\begin{align*}
	\z := 
	\col(\g, \es).
\end{align*}
With Lemma~\ref{lemma:ss} at hand, let us introduce the error coordinates $\tc \in \R^{\nc}$ and $\tz \in \R^{\nz}$ defined as
\begin{align}
	\label{eq:error_coordinates}
	\begin{bmatrix}
		\w
		\\
		\chi
		\\
		\z
	\end{bmatrix}
	 \longmapsto
	\begin{bmatrix}
		\w
		\\
		\tc 
		\\
		\tz
	\end{bmatrix}  := 
	\begin{bmatrix} 
		\w 
		\\
		\begin{bmatrix}
			I_{\dimx}& 0
			\\
			0& \ssz I_{(\dimx+\dimu)(2\dimx+\dimu)}
		\end{bmatrix}\left(\chi  -  \sts(\w)\right)
		\\
		\z - 
		\begin{bmatrix}
			\gstar
			\\
			\pr	
		\end{bmatrix}
	\end{bmatrix}.
\end{align}
For notational convenience, we will sometimes refer to the components of $\tz$ as $\col(\tg,\tes)$.
Finally, the closed-loop dynamics~\eqref{eq:algo_closed_loop} in the new coordinates~\eqref{eq:error_coordinates} reads as
\begin{subequations}\label{eq:mixed_scale_system_our}
	\begin{align}
		\tctp &= \cA(\tzt)\tct + h(\tzt,\iter) + \ssz g(\tct,\tzt,\iter)\label{eq:mixed_scale_system_our_fast}
		\\
		\tztp &= \tzt + \ssz f(\tct,\tzt, \iter),\label{eq:mixed_scale_system_our_slow}
	\end{align}
\end{subequations}
where $\cA(\tz) := \cAc(\tz_1 + \gstar)$ (cf.~\eqref{eq:cAc})
and we introduced  
$h: \R^{\nz} \times \N \to \R^{\nc}$, $g: \R^{\nc} \times \R^{\nz} \times \N \to \R^{\nc}$, and $f: \R^{\nc} \times \R^{\nz} \times \N \to \R^{\nz}$ defined respectively as 
\begin{subequations}\label{eq:h_g_f}
	\begin{align}
		h(\tz,\iter) 
		& := 
		\begin{bmatrix}
			\Bs\tz_1\pix\wt
			\\
			0_{(\dimx+\dimu)(2\dimx+\dimu)}		
		\end{bmatrix}
		\\
		g(\tc,\tz,\iter)&
		 :=  
		\begin{bmatrix} 
			0_{\dimx}
			\\
			 \fun_2(\tc   +   \sts(\wt),\tz_1  +   \gstar,\wt)  -   \fun_2(\sts(\wt),\gstar,\wt)
		\end{bmatrix}
		\\
		f(\tc,\tz,\iter)
		&:= 
		\begin{bmatrix}
			f_1(\tc,\tz,\iter) \\
			f_2(\tc,\tz,\iter) 
		\end{bmatrix}
		,
		\label{eq:f}
	\end{align}
	with
	\begin{align}\label{eq:f_1}
		f_1(\tc,\tz,\iter) & :=
		-\G(\tz_1 + \gstar,\tz_2 + \pr)
		\\
		\nonumber
		f_2(\tc,\tz,\iter) & := 
		-(\unvc{\tc_2} + \hssst)^\dagger
		\Big((\unvc{\tc_2} + \hssst)\tz_2
		\\
		& 
		\qquad 
		\qquad 
		+ \unvc{\tc_2 - \tc_3}\pr\Big),
		\label{eq:f_2}
	\end{align}
\end{subequations}
where, for the sake of readability, in~\eqref{eq:h_g_f} we used the shorthands $\tc := \col(\tc_1,\tc_2,\tc_3)$ and $\tz = \col(\tz_1,\tz_2)$, 
we used the partition $\fun(\chi,\g,\w) := \col(\fun_1(\chi,\g,\w),\fun_2(\chi,\g,\w))$ (cf.~\eqref{eq:phi}), 
we defined $\hssst \in \R^{(n+m)\times(n+m)}$ as
\begin{align}
	\hssst &:=\opH(\pih, \wt),\label{eq:hssst}
\end{align}
which represents the steady-state value of $\hst$ (see~\eqref{eq:v_h} for the definition of $\opH$)
and we introduced the error coordinates $\ths \in \R^{(n+m)\times(n+m)}$ and $\trr \in \R^{(n+m)\times n}$, as

\begin{align}
	\begin{bmatrix}
		\w
		\\
		\vhs
		\\
		\vrr
	\end{bmatrix}
	\!\! \longmapsto	\!\!
	\begin{bmatrix}
		\w
		\\
		\ths
		\\
		\trr
	\end{bmatrix}
	\!\! := \!\!
	\begin{bmatrix}
		\w
		\\
		\unvc{\vhs - \opH(\pih, \w)}
		\\
		\unvc{\vrr \! - \! \pir\vc{\w\w\T}}
	\end{bmatrix} \!\!. \!\!
	\label{eq:ths}
\end{align}%
We point out that with this transformation we obtained a dynamical system with two-time scales as the one described in Section~\ref{app:averaging} (cf.~system~\eqref{eq:mixed_scale_system}).
As customary in this context, we distinguish between (i) the fast dynamics~\eqref{eq:mixed_scale_system_our_fast} with state $\tc$, and (ii) the slow one~\eqref{eq:mixed_scale_system_our_slow} with state $\tz$. 
Fig.~\ref{fig:block_diagram} shows the mentioned interconnected structure of system~\eqref{eq:mixed_scale_system_our}.
In this reformulation, the effect of the exogenous signal $\wt$ is embedded in the time dependency of $h$, $g$, and $f$. %
Finally, by definition of $h$, $g$, and $f$ (cf.~\eqref{eq:h_g_f}) and since $\G(\gstar,\pr) = 0$, %
for all $\iter \in \N$, we have
\begin{align}
	h(0,\iter) = 0, \qquad
	g(0,0,\iter) = 0,  \qquad
	f(0,0,\iter) = 0.\label{eq:equilibrium}
\end{align}

\subsection{Averaged System Analysis}
\label{sec:average_system}

Next, we carry out the stability analysis of the time-varying system~\eqref{eq:mixed_scale_system_our} by using the averaging and timescale separation theories (cf. Section~\ref{app:averaging}).
System~\eqref{eq:mixed_scale_system_our} enjoys a two-time-scale structure.
Hence, we can study~\eqref{eq:mixed_scale_system_our} by only investigating an auxiliary system typically termed the \emph{averaged system} (see Section~\ref{app:averaging}). 
The latter is obtained by considering the slow dynamics~\eqref{eq:mixed_scale_system_our_slow} in which the fast state is frozen to its equilibrium ($\tct = 0$ for all $\iter \in \N$) and the vector field describing the dynamics is averaged with respect to time.
The following result is instrumental to properly write the averaged system.
\begin{lemma}\label{lemma:f_av}
	Let the assumptions of Theorem~\ref{th:convergence} hold true.
	Consider $f$ defined in~\eqref{eq:f}.
	Then, it holds  
	\begin{align}\label{eq:fav}
		\lim_{T \to \infty}\frac{1}{T} \sum_{\initer = \bar{\iter} + 1}^{\bar{\iter} + T}f(0,\tz,\initer) = \! - \!
		\begin{bmatrix}
			\G(\tg \! + \! \gstar,\tes \! + \! \pr)
			\\
			\tes
		\end{bmatrix}
	\end{align}
	uniformly in $\bar{\iter} \in \N$ and for all $\tz = \col(\tg,\tes) \in \R^{\nz}$.\oprocend
\end{lemma}
The proof of Lemma~\ref{lemma:f_av} is given in Appendix~\ref{app:f_av}.

Lemma~\ref{lemma:f_av} provides a suitable approximation of the dynamics of $\tz$ in~\eqref{eq:mixed_scale_system_our_slow} when (i) the convergence of the fast state $\tc$ to its equilibrium has already occurred and (ii) by averaging over time $\iter$ the vector field $f(0,\tz,\iter)$.
Specifically, under this approximation, Lemma~\ref{lemma:f_av} ensures that the two components of the driving term of the dynamics of $\tz$ are given by (i) a proportional term $-\ssz\tes$ and (ii) an approximated version of the correct gradient $\G(\tg + \gstar, \pr)$.
Next, we will leverage averaging theory to prove the stability of the origin for system~\eqref{eq:mixed_scale_system_our}.

\begin{figure}[!htpb]
	\centering
	\includegraphics[scale=\scalebd]{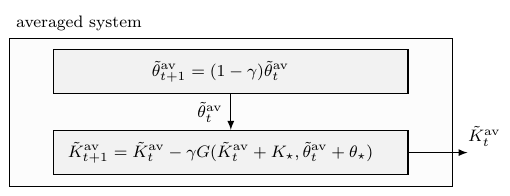}
	\caption{Block diagram of~\eqref{eq:algo_closed_loop_average} with $\tzat = \col(\tgat,\tesat)$.
	\label{fig:cascade_average}}
\end{figure}

Once the averaged vector field has been characterized in Lemma~\ref{lemma:f_av}, we can introduce $f\av: \R^{\nz} \to \R^{\nz}$ given by
\begin{align*}
	f\av(\tz) := \lim_{T \to \infty}\tfrac{1}{T}\textstyle\sum_{\initer = \bar{\iter} + 1}^{\bar{\iter} + T} f(0,\tz,\iter).
\end{align*}%
Then, the averaged system associated to~\eqref{eq:mixed_scale_system_our} reads as
\begin{align}
	\tzatp = \tzat + \ssz f\av(\tzat),
	\label{eq:algo_closed_loop_average}
\end{align}
with state $\tzat : =\col(\tgat,\tesat) \in \R^{\nz}$.
Expanding the expression of $f\av$ (cf.~\eqref{eq:fav}), the dynamics in~\eqref{eq:algo_closed_loop_average} results in a cascade as depicted in Fig.~\ref{fig:cascade_average}.
For the sake of compactness, we also introduce the (averaged) estimates $A\av\ud\iter \in \R^{\dimx \times \dimx}$ and $B\av\ud\iter \in \R^{\dimx \times \dimu}$ of $A$ and $B$, defined as
\begin{align}\label{eq:AavBav_thetaav}
	[A\av\ud\iter 
		\;
		B\av\ud\iter]\T := \tesat + \pr.
\end{align}
We establish exponential stability of the origin for~\eqref{eq:algo_closed_loop_average}.
\begin{proposition}\label{prop:average}
	Let the assumptions of Theorem~\ref{th:convergence} hold true.
	Consider the averaged system~\eqref{eq:algo_closed_loop_average}. %
	Then, there exists $\bar{\ssz}\av > 0$ such that, for all $\ssz \in (0,\bar{\ssz}\av)$, the origin of~\eqref{eq:algo_closed_loop_average} is exponentially stable. 
	\oprocend
\end{proposition}
The proof of Proposition~\ref{prop:average} is given in Appendix~\ref{app:proof}.

Once this result has been posed, we can proceed with the proof of Theorem~\ref{th:convergence} in the next subsection.

\subsection{Proof of Theorem~\ref{th:convergence}}
\label{sec:proof_of_convergence}
	
We will use Theorem~\ref{th:bai} given in Section~\ref{app:averaging} to guarantee the exponential stability of the origin for~\eqref{eq:mixed_scale_system_our}.
Specifically, in order to apply Theorem~\ref{th:bai}, we need to verify 
\begin{enumerate}[(i)]
	\item the exponential stability of the origin for the averaged system~\eqref{eq:algo_closed_loop_average};
	\item the Lipschitz continuity of the vector field of the original system~\eqref{eq:mixed_scale_system_our} (cf. Assumption~\ref{ass:lipschitz});
	\item that the origin is an equilibrium point of the original system~\eqref{eq:mixed_scale_system_our} (cf. Assumption~\ref{ass:equilibrium});
	\item that the matrix function $\cA(\tz)$ satisfies Assumption~\ref{ass:schur};
	\item that the difference between the vector fields $f$ and $f\av$ of the original~\eqref{eq:mixed_scale_system_our_slow} and~\eqref{eq:algo_closed_loop_average}, respectively, satisfies
\begin{subequations}\label{eq:conditons_tilde_f}
	\begin{align}
		&\norm{\tfrac{1}{T}\textstyle\sum_{\initer = \bar{\iter} + 1}^{\bar{\iter} + T}\Delta f(\tz,\initer)} \leq \nu(T)\norm{\tz}
		\\
		&\norm{\tfrac{1}{T}\textstyle\sum_{\initer = \bar{\iter} + 1}^{\bar{\iter} + T}\frac{\partial\Delta f(\tz,\initer)}{\partial \tz}} \leq \nu(T),
	\end{align}
\end{subequations}
for all $\initer \in \N$, where $\Delta f(\tz,\initer) := f(0,\tz,\initer) - f\av(\tz)$ and $\nu(\iter)$ is a nonnegative strictly decreasing function with the property $\nu(\iter) \to 0$ as $\iter \to \infty$ (see Assumption~\ref{ass:nu}).
\end{enumerate}

Condition (i) follows from Proposition~\ref{prop:average}.
Condition (ii) is satisfied by using the quantities in~\eqref{eq:lipp} in Appendix~\ref{app:proof} and the invertibility of $\hssst$ (cf. Lemma~\ref{lemma:inv} in Appendix~\ref{app:proof}) to find the required Lipschitz constants of the vector field of~\eqref{eq:mixed_scale_system_our}.
Condition (iii) is verified by~\eqref{eq:equilibrium}.
As for condition (iv), we note that $\cA(\tz)$ is Schur for all $\tg \in \R^{m \times n}$ (the first component of $\tz$, see its definition in~\eqref{eq:error_coordinates}) such that $(\tg + \gstar) \in \cK$ by definition of $\cK$.
Hence, condition (iv) is verified with the largest ball contained in $\cK$ and centered in $\gstar$.
Finally, to check condition (v) (cf.~\eqref{eq:conditons_tilde_f}), we use the definitions of $f$ (cf.~\eqref{eq:f}) and $f\av$ (cf.~\eqref{eq:fav}) to write
\begin{align}
	\Delta f(\tz,\iter) = \begin{bmatrix}
		0
		\\
		(\hssst)^\dagger\hssst\tes- \tes
	\end{bmatrix} \stackrel{(a)}{=} 0,
	\label{eq:f_tilde}
\end{align}
where in $(a)$ we used the fact that $\hssst$ is actually a square invertible matrix for all $\iter \in \N$ (cf. Lemma~\ref{lemma:inv} in Appendix~\ref{app:proof}). 
Therefore, the conditions in~\eqref{eq:conditons_tilde_f} are satisfied and, thus, the proof follows by Theorem~\ref{th:bai}.
\section{Numerical Simulations}
\label{sec:numerical_simulations}

In this section, we numerically test our strategy.
We consider the model of the longitudinal dynamics of a highly maneuverable aircraft linearized at an altitude of $3000$ $[$ft$]$ and a velocity of $0.6$ $[$Mach$]$, see~\cite{kapasouris1990design}.
The resulting linear continuous-time time-invariant dynamics reads as 
\begin{align}
	\label{eq:f16:ct}
	\begin{split}
		\dot{x} =& 
		\left[\begin{smallmatrix}
			 -0.0151 & -60.5651 & 0 & -32.174
			 \\
	     -0.0001 & -1.3411& 0.9929& 0
			 \\
	      0.00018 & 43.2541 & -0.86939 & 0
				\\
	      0 & 0 & 1 & 0
		\end{smallmatrix}\right]
		x 
		+
		\left[\begin{smallmatrix}
			-2.516  &-13.136
			\\
	     -0.1689  & -0.2514
			 \\
	     -17.251 & -1.5766
			 \\
	      0 & 0
		\end{smallmatrix}\right]
		u,
	\end{split}
\end{align}
where $x \in \R^4$ contains the forward velocity, the attack angle, the pitch rate and the pitch angle, while $u \in \R^2$ contains the elevator and flaperon angles.
The discrete-time system matrices $\As$ and $\Bs$ are obtained from the continuous-time ones in~\eqref{eq:f16:ct} with a Forward-Euler discretization with sampling time $T_s = 0.05$ $[$s$]$. 
The cost matrices $Q \in \R^{4 \times 4}$ and $R \in \R^{2 \times 2}$ are randomly generated ensuring that $Q = Q\T \ge 0$ and $R = R\T > 0$.
As for the design of $F$ and $E$, we use the procedure outlined in Remark~\ref{rem:F_E} by choosing $\omega_1 = 1/5$, $\omega_{i+1} = \omega_{i}$ if $i$ is odd and $\omega_i = 2\omega_{i-2}$ if $i$ is even for all $i \in \{2,\dots,\nw\}$.
Further, we set $\ssz = 5\cdot10^{-4}$ and $\w\ud 0 = [0.01\; 0\; \dots\; 0.01\; 0]\T$.
Finally, the initial condition $\x\ud0$ is sampled from a normal distribution with mean value $10$ for each state.

\subsection{Aircraft Control}
\begin{figure}[htpb]
\centering
\includegraphics[scale = \scalefig]{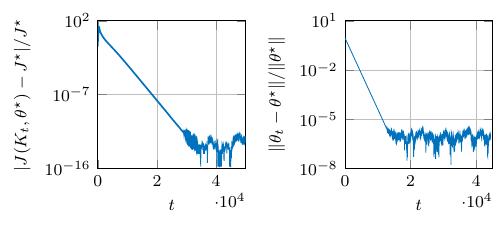}
\caption{Evolution of the (left) cost error $|J(\gt,\est) - J^\star|/J^\star$ and (right) estimation error $\norm{\est - \pr}/\norm{\pr}$.}
\label{fig:lti:state_traj}
\end{figure}

\begin{figure}[htpb]
\centering
\includegraphics[scale = \scalefigzoom]{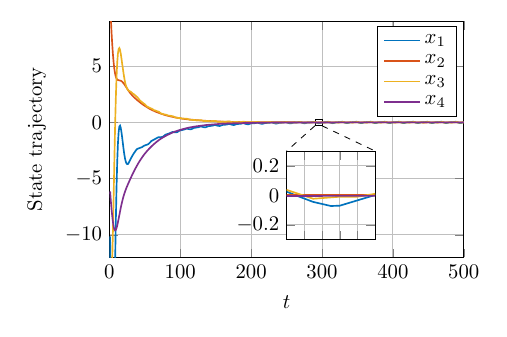}
\caption{State trajectory of the closed-loop system.} 
\label{fig:lti:results}
\end{figure}

We consider system~\eqref{eq:f16:ct} and run \algname/
with the exogenous signal generated via the procedure detailed above.
Fig.~\ref{fig:lti:state_traj} shows the evolution of the normalized (left) cost error $|J(\gt,\est) - J^\star|/J^\star$ (with $J^\star := J(\gstar,\pr)$ and $\pr := [\As\; \Bs]\T$) and (right) estimation error $\norm{\est - \pr}/\norm{\pr}$.
The convergence to the optimal cost $J^\star$ and true parameters $\pr$ is achieved.
Finally, Fig.~\ref{fig:lti:results} shows the closed-loop system trajectory. 
After a transient, the states oscillate about the origin due to $\drt$.

\subsection{Aircraft Control with Drifting Parameters}

\begin{figure}[htpb]
	\centering
	\includegraphics[scale = \scalefig]{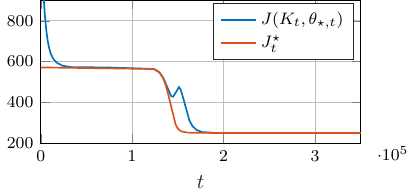}
	\caption{Comparison between $\J(\gt,\prt)$ and $\J^\star\ud\iter$.}
	\label{fig:aircraft_comparison}
\end{figure}

\begin{figure}[htpb]
\centering
\includegraphics[scale = \scalefig]{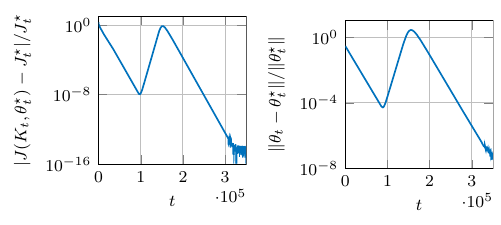}
\caption{Evolution of the (left) cost error $|J(\gt,\prt) - J^\star\ud\iter|/J^\star\ud\iter$ and (right) estimation error $\norm{\est - \prt}/\norm{\prt}$.}
\label{fig:results_variations}
\end{figure}

To better highlight the capabilities of our algorithm, we also consider the case where 
the system matrices $\As$, $\Bs$, slowly change over time. The new time-varying state and input matrices are denoted as $\Asv$ and $\Bsv$, respectively.
More in detail, 
the time-varying system matrices $\Asv$ and $\Bsv$ smoothly evolve from $\As$ and $\Bs$ toward a new pair of matrices $A_+$ and $B_+$, according to the update law 
\begin{align*}
	\begin{bmatrix}
		\Asv
		&
		\Bsv
	\end{bmatrix} &= (1 - \sigma(\iter))\begin{bmatrix}
		\As
		&
		\Bs
	\end{bmatrix} + \sigma(\iter)
	\begin{bmatrix}
		A_+
		&
		B_+
	\end{bmatrix},
\end{align*}
for all $\iter \!\in\! \N$, with $\sigma(\iter)$ being the sigmoid function $\sigma(\iter) = 1/(1 + \exp((\iter - \tr)/\alpha\big))$,
where $\alpha \in \R$ determines the transition width and $\tr \in \N$ defines the transition center.
We select $\tr \! =  \! 1.5\cdot10^5$, $\alpha \! = \! 5\cdot10^3$, while the entries $\theta_{ij}^{+}$ of $[A_+\; B_+] \in \R^{n \times \n m}$ are randomly generated as
\begin{align*}
	\theta_{ij}^{+} &= \begin{cases}
		\theta_{ij} \quad \quad\quad\quad \text{if }\theta_{ij} = 0
		\\
		\theta_{ij} + \var \theta_{ij}^A \quad \text{otherwise},
	\end{cases}
\end{align*}
for all $i \in \until{\n}$ and $j \in \until{\n m}$, where $\theta_{ij}^A$ is a random variable normally distributed and $\var = 0.1$ is a scale factor.
Fig.~\ref{fig:aircraft_comparison} compares $\J(\gt,\prt)$ and $\J^\star\ud\iter:= J(\gstart,\prt)$, where $\prt := [\Asv\; \Bsv]\T$ and $\gstart$ is the corresponding optimal gain.
Finally, Fig.~\ref{fig:results_variations} shows the evolution of the normalized (left) cost error $|\J(\gt,\prt) - \J^\star\ud\iter|/\J^\star\ud\iter$ and (right) estimation error $\norm{\est - \prt}/\norm{\prt}$.
In both cases, asymptotic tracking of $J^\star\ud\iter$ and $\prt$ is achieved.
As one may expect, in the neighborhood of the inflection point $\iter\approx\tr$, both errors increase.
However, we note the adaptability of our policy as it quickly recovers convergence toward $J^\star\ud\iter$ and $\prt$.

\section{Conclusions}

In this paper, we addressed infinite-horizon LQR problems with unknown system matrices.
We proposed a method mixing the identification of the unknown matrices with the optimization of the feedback policy.
We proved exponential convergence of the overall closed-loop system to the optimal steady-state associated to the optimal gain and the exact matrices by using tools from Lyapunov-based analysis and averaging theory.
Although our analysis considers a time-invariant plant without disturbances, our procedure is inherently applicable to more challenging scenarios in which, e.g., the system and cost matrices vary over time and disturbances affect the plant.
Thus, our work paves the way for stability certificates also in these more complex settings.
Future research will focus on extending the proposed framework to stochastic settings and on studying the connections between regret and stability analysis.

\appendix
\numberwithin{equation}{section}

\renewcommand{\thesection}{\Alph{section}}

\section{Proof of Lemma~\ref{lemma:ss}}
\label{app:ss}

Since~\eqref{eq:steady_state_locus} is obtained by setting $\gt = \gstar$ in~\eqref{eq:compact_system} (which collects~\eqref{eq:algo_closed_loop_drt},~\eqref{eq:algo_closed_loop_xt}, and~\eqref{eq:algo_closed_loop_with_vc}), we study~\eqref{eq:algo_closed_loop_drt}-\eqref{eq:algo_closed_loop_xt} in the manifold in which $\gt = \gstar$, namely
\begin{subequations}\label{eq:x_dynamics_restricted}
\begin{align}
	\wtp &= \Sw\wt
	\\
	\xtp &= (\As + \Bs\gstar)\xt + \Bs\EE\wt.
\end{align}
\end{subequations}
The steady-state locus of the cascade system~\eqref{eq:x_dynamics_restricted} is $\col(\wt,\xt) = \col(I_{\nw},\pix)\wt$, with $\pix \in \R^{n \times \nw}$ such that 
\begin{align}\label{eq:pix_equation}
	\pix \Sw = (\As + \Bs\gstar)\pix + \Bs\EE.
\end{align}
Since $\sigma(\Sw) \cap \sigma(\As + \Bs\gstar) = \emptyset$, $\pix$ exists and is unique.
Then, we study~\eqref{eq:algo_closed_loop_with_vc} restricted to the manifold in which $(\xt,\gt) = (\pix\wt, \gstar)$. %
Let $M \in \R^{(n+m)\times(n+m)}$ be
\begin{align}
	M := 
	\begin{bmatrix}
		\pix\T
		&
		(\gstar\pix + E)\T
	\end{bmatrix}\T,
	\label{eq:M}
\end{align}
then it holds
\begin{subequations}\label{eq:vectorized_equations}
\begin{align}
	&\vc{\wtp\wtp\T} = \vc{\Sw\wt\wt\T \Sw\T} 
	\label{eq:vectorized_equations_1}
	\\
	&\vhstp = \lambda \vhst
	+ \vc{M\wt\wt\T M\T}
	\\
	&\vrrtp = \lambda \vrrt
	+ \vc{M\wt\wt\T M\T\pr},
\end{align}
\end{subequations}
where~\eqref{eq:vectorized_equations_1} comes from the vectorization of~\eqref{eq:algo_closed_loop_drt}. %
By using the vectorization properties\footnote{Given any $X_1 \in \R^{n_1 \times n_2}$, $X_2 \in \R^{n_2 \times n_3}$, and $X_3 \in \R^{n_3 \times n_4}$, it holds $\vc{X_1X_2X_3} = (X_3\T \otimes X_1)\vc{X_2}$.}, we rewrite~\eqref{eq:vectorized_equations} as
\begin{subequations}\label{eq:vectorized_equations_2}
	\begin{align}
		&\hspace{-.2cm}\svc{\wtp\wtp\T} \!=\! (\Sw \otimes \Sw) \vc{\wt\wt\T}
		\\
		&\vhstp \!=\! \lambda\vhst 
		\!+\! (M \otimes M)\vc{\wt\wt\T}\label{eq:vec_hs_update}
		\\
		&\vrrtp \!=\!  \lambda \vrrt 
		\!+\! ({\pr}\T\!  M \! \otimes \! M)\vc{\wt\wt\T}\!.
		\label{eq:vec_rr_update}
	\end{align}
\end{subequations}
We note that~\eqref{eq:vectorized_equations_2} describes a cascade of linear systems with states $\vc{\wt\wt\T}$ and $(\vhst,\vrrt)$.
Hence, it is well known that its steady-state can be characterized by resorting to a Sylvester equation.
To this end, let $\pih \in \R^{(n+m)^2 \times \nw^2}$ and $\pir \in \R^{(n+m)n \times \nw^2}$ solve
\begin{subequations}\label{eq:sylvester_vec}
	\begin{align}
		\pih(\Sw \otimes \Sw) &= \lambda\pih + M \otimes M\label{eq:sylvester_vec_H}
		\\
		\pir(\Sw \otimes \Sw) &= \lambda\pir + ({\pr}\T M) \otimes M.\label{eq:sylvester_vec_rr}
	\end{align}
\end{subequations}
Since $\sigma(\Sw \otimes \Sw) \cap \sigma(\lambda I) = \emptyset$, $\pir$ and $\pih$ exist and are unique~\cite{bhatia1997and}.
The proof of~\eqref{eq:steady_state_locus} follows by the definition of $\sts$ (cf.~\eqref{eq:ss}) and plugging~\eqref{eq:pix_equation} and~\eqref{eq:sylvester_vec} into~\eqref{eq:compact_system}.

To show~\eqref{eq:pih_pir_pr}, 
we multiply~\eqref{eq:sylvester_vec_H} by ${\pr}\T \otimes I_{n + m}$ and get %
\begin{align}
	({\pr}\T \! \otimes \! I_{n + m})\pih(\Sw \otimes \Sw \!-\! \lambda I) 
	&
	\! = \! ({\pr}\T \otimes I_{n + m})(\M \otimes \M)
	\notag\\
	&\! \stackrel{(a)}{=} \!
	({\pr}\T \M) \otimes \M
	\notag\\
	&\! \stackrel{(b)}{=} \!
	\pir(\! \Sw \! \otimes \! \Sw \! - \! \lambda I),
\end{align}
where $(a)$ uses a property of the Kronecker operator\footnote{Given any $X_1 \! \in \! \R^{n_1 \times n_2}$, $X_2 \! \in \! \R^{n_3 \times n_4}$, $X_3 \!\in\! \R^{n_2 \times n_5}$, and $X_4 \!\in\! \R^{n_4 \times n_6}$, it holds $(X_1 \! \otimes \! X_2)(X_3 \! \otimes\! X_4) \! = \!  (X_1X_3) \! \otimes \! (X_2X_4)$.}, while $(b)$ follows from~\eqref{eq:sylvester_vec_rr}. The proof is complete.

\section{Proof of Lemma~\ref{lemma:f_av}}
\label{app:f_av}

To prove Lemma~\ref{lemma:f_av}, we need the following result.
\begin{lemma}\label{lemma:inv}
	Let the assumptions of Theorem~\ref{th:convergence} hold true.
	Then, $\hssst$ (cf.~\eqref{eq:hssst}) is invertible for all $\iter\in \N$.
\end{lemma}%
\begin{proof}
We prove the invertibility of $\hssst$ by studying the evolution of $\hst$.
Its dynamics~\eqref{eq:algo_closed_loop_hst} restricted to the manifold in which $\xt = \pix\wt$ and $\gt = \gstar$ reads as
\begin{align}\label{eq:hs_manifold}
	\hstp = \lambda\hst + M\wt\wt\T M\T,
\end{align}
with $M$ as in~\eqref{eq:M}.
The explicit solution of~\eqref{eq:hs_manifold} is
\begin{align}\label{eq:hs_closed_form}
	\hst = \lambda^{\iter}\hs\ud 0 + M \underbrace{\left(\sum_{\initer = 0}^{\iter-1}\lambda^{\iter -1 - \initer}\w\ud\initer\w\ud\initer\T \right)}_{=:\Dt}M\T.
\end{align}
Since $\lambda \in (0,1)$, the free evolution $\lambda^\iter \hs\ud0$ of~\eqref{eq:hs_closed_form} asymptotically vanishes and does not affect the invertibility of $\hst$.
Thus, we focus on the forced response $M \Dt M\T$ only. 
Since $\alpha_1 I_{\nw} \le \sum_{\initer=\bar{\iter}+1}^{\bar{\iter} + \per}\w\ud \initer \w\ud\initer\T \le \alpha_2 I_{\nw}$
for all $\bar{\iter} \in \N$ (cf. Assumption~\ref{ass:dither}), we have $\Dt > 0$ for all $\iter \ge \per$ by~\cite[Lemma~1]{johnstone1982exponential}.
Let us consider the Cholesky decomposition of $\Dt$ given by $\Dt = \Ct\Ct\T$, with $\Ct \in \R^{\nw \times \nw}$ invertible.
Then\footnote{%
$\rank(X X\T) \! = \! \rank(X) \! =  \! \rank(X\T)$ for all $X \in \R^{n \times m}$.}, for all $\iter \ge \per$, we can write
\begin{align}
	\rank\left( M \Dt M\T \right) 
	&= 
	\rank(M\Ct) 
	\stackrel{(a)}{=} \rank(M),
	\label{eq:rank_equality_intermediate}
\end{align}
where $(a)$ uses full-rankness of $\Ct$ and a rank property.\footnote{Given $X_1 \in \R^{n_1 \times n_2}$ and $X_2 \in \R^{n_2 \times n_3}$ it holds $\rank(X_1X_2) = \rank(X_1)$ if $\rank(X_2) = n_2$.}
To compute $\rank(M)$, we study system~\eqref{eq:algo_closed_loop_drt}-\eqref{eq:algo_closed_loop_xt} restricted to the manifold in which $\gt = \gstar$, namely
\begin{subequations}\label{eq:x_dynamics_restricted_two}
	\begin{align}
		\wtp &= \Sw\wt
		\\
		\xtp &= (\As + \Bs\gstar)\xt + \Bs\EE\wt.
		\label{eq:x_dynamics_restricted_two_xt}
	\end{align}
\end{subequations}
Recalling that $\drt = \EE\wt$ satisfies condition~\eqref{eq:PE_order} (cf. Assumption~\ref{ass:dither}) and that $(\As,\Bs)$ is controllable (cf. Assumption~\ref{ass:unknown}), we can invoke~\cite[Cor.~2]{willems2005note} to claim that
\begin{align}
	\rank\left(\begin{bmatrix}
		\x\ud0& \dots& \x\ud{\tPE-1}
		\\
		\dr\ud0& \dots& \dr\ud{\tPE-1}
	\end{bmatrix}\right) = n + m,
	\label{eq:X_U}
\end{align}
for all $(\x\ud 0, \dr\ud 0) \in \R^{n} \times \R^{m}$.
When the initial condition of~\eqref{eq:x_dynamics_restricted_two_xt} lies in the steady-state locus (cf.~\eqref{eq:pix_equation}) (i.e., $\x\ud0 = \pix\w\ud0$), the condition in~\eqref{eq:X_U} simplifies to
\begin{align}
	\rank\left(M\begin{bmatrix}
		\w\ud0& \dots& \w\ud{\tPE-1}
	\end{bmatrix}\right) 
	=
	n + m,
	\label{eq:X_U_M}
\end{align}
which implies\footnote{Given $X_1 \in \R^{n_1 \times n_2}$ and $X_2 \in \R^{n_2 \times n_3}$, it holds $\rank(X_1X_2) \leq \min\{\rank(X_1),\rank(X_2)\}$.} 
	$\rank(M) \ge n + m$. %
Being $\rank(M) \leq n + m$ by construction, the last inequality yields $\rank(M) = n + m$, which combined with~\eqref{eq:rank_equality_intermediate}, leads to
\begin{align}
	\rank\left( M \Dt M\T \right) = n + m,
	\label{eq:rank_equality_1}
\end{align}
for all $\iter \ge \per$.
To study $\rank( M \Dt M\T )$, we note that, since
$\lambda \in (0,1)$, $M \Dt M\T$ exponentially converges to $\hssst$.
Thus, by continuity, there exists $\iter_{\infty} \ge \per$ such that
\begin{align*}
	\rank(\hssst) = n + m,
\end{align*} 
for all $\iter \ge \iter_{\infty}$.
Finally, being $\hssst$ a static function of the periodic signal $\wt$, then $\hssst$ is periodic as well so that its full-rankness is independent of $\iter$. Thus, it must be that $\rank(\hssst) = n + m$ for all $\iter \in \N$, and the proof follows.
\oprocend
\end{proof}
Now, let us label the two components of $f\av$ as
\begin{align}\notag
	\begin{bmatrix}
		f\av_1(\tz)\T
		&
		f\av_2(\tz)\T
	\end{bmatrix}\T := \lim_{T \to \infty}\tfrac{1}{T} \textstyle\sum_{\initer = \bar{\iter} + 1}^{\bar{\iter} + T}f(0,\tz,\initer).
\end{align}
The first block $f\av_1(\tz)$ of $f$ (cf.~\eqref{eq:f_1}) does not depend on $\iter$ and, thus, it trivially coincides with $f_1$, namely
\begin{align} \notag %
	f\av_1(\tz) := -\G(\tg + \gstar, \tes + \pr),
\end{align}
with $\tz = \col(\tg,\tes)$.
As for $f\av_2(\tz)$, with $\tc = 0$, it holds
\begin{align}\notag %
	f\av_2(\tz) 
	=
	 -\lim_{T \to \infty}\tfrac{1}{T}\textstyle\sum_{\initer = \bar{\iter} + 1}^{\bar{\iter} + T }(\hsss\ud{\initer})^\dagger\hsss\ud{\initer}\tes 
	\stackrel{(a)}{=}
	-\tes,
\end{align}
where the last equality uses Lemma~\ref{lemma:inv}. %

\section{Proof of Proposition~\ref{prop:average}}
\label{app:proof}

The proof resorts to a suitable Lyapunov candidate function whose increment along trajectories of system~\eqref{eq:algo_closed_loop_average} will allow us to claim exponential stability of the origin.

To ease the notation, we use $\tza := \col(\tga,\tesa)$. 
By~\cite[Lemma~3.12]{bu2019lqr}, $\J$ is gradient dominated, namely, it holds
\begin{align}
	\J(\g,\pr)  -  \J(\gstar,\pr) \leq \gd \norm{\G(\g,\pr)}^2,\label{eq:gradient_dominance}
\end{align}
for all $\g \in \cK$ and some $\gd > 0$. %
We now consider the Lyapunov function $\Vav: \cK \times \dpr \to \R$ defined as 
\begin{align}
	& \Vav(\tga,\tesa) \! := \! \label{eq:Vav}
	\mav(\J(\tga \! + \! \gstar,\pr) \! - \! \J(\gstar,\pr))
	\! + \! \halff \|\tesa\|^2,
\end{align}
where $\mav > 0$ will be set later.
Since $(\As,\Bs)$ is controllable (cf. Assumption~\ref{ass:unknown}) and $\As + \Bs\gstar$ is Schur, by continuity (see, e.g.,~\cite[Thm.~10]{klamka2008controllability} and~\cite[Thm.~6.3.12]{horn2012matrix}), there exists a neighborhood, say it $\set \subseteq  \cK \times \dpr$, of $(\gstar,\pr)$ such that (i) $(A,B)$ is controllable and (ii) $A + B\g$ is Schur for every pair $(\g,\es) := (\g,[A\; B]\T) \in \set$.
Let $\rmax > 0$ be the radius of the largest ball centered in $(\gstar,\pr)$ and contained in $\set$. 
Then, let us arbitrarily choose radius $\tilde{r}\av \in (0,\rmax)$ and use it to define the constants %
\begin{subequations}\label{eq:lipp}
	\begin{align}
		\lipp_0 \! &:= \!\!\!\!\max_{(\tga,\tesa) \in \cB_{\tilde{r}\av}(0_{\nz})}\!\!\{ \snorm{\G(\tga \! + \!  \gstar,\tesa  \! + \!  \pr)}\}
		\!\!
		\label{eq:lipp0}
		\\
		\lipp_1 \! &:= \!\! \max_{(\tga,\tesa) \in \cB_{\tilde{r}\av}(0_{\nz})}\!\!\left\{ \norm{\tfrac{\partial \G(\tga \! + \!  \gstar,\tesa  \! + \!  \pr)}{\partial \tga}}\right\}
		\label{eq:lipp1}
		\\
		\lipp_2 \! &:= \!\! \max_{(\tga,\tesa) \in \cB_{\tilde{r}\av}(0_{\nz})}\!\!\left\{ \norm{\tfrac{\partial \G(\tga \! + \! \gstar, \tesa \! + \! \pr)}{\partial \tesa}}\right\}\!.\label{eq:lipp2}
	\end{align}
\end{subequations}
Then, we use the decomposition $\tesa := [\tesa_A\; \tesa_B]\T$ to write
\begin{align}
	&(\tga,\tesa) \in \cB_{\tilde{r}\av}(0_{\nz})
		\implies
		 (\tesa_{A}  + \As)+ (\tesa_{B} + \Bs) (\tga +\gstar) \text{ is Schur}.\!\!
		 \label{eq:implications}
\end{align}
By combining~\cite[Proposition~3.10]{bu2019lqr} (and similar arguments) with~\eqref{eq:implications}, it can be shown that the functions $\G(\tga+\gstar,\tesa+\pr)$, $\partial \G(\cdot + \gstar,\cdot + \pr)/\partial \tga$, and $\partial \G(\cdot + \gstar, \cdot + \pr)/\partial \tesa$ are continuous in $\cB_{\tilde{r}\av}(0_{\nz})$.
In turn, since $\cB_{\tilde{r}\av}(0_{\nz})$ is compact by definition, this implies that $\lipp_0$, $\lipp_1$, and $\lipp_2$ are finite.
With this result at hand, we show that, for a proper bound on $\ssz$, the set $\cB_{r\av}(0_{\nz})$ is forward invariant for~\eqref{eq:algo_closed_loop_average} for some $r\av \in (0,\tilde{r}\av]$ that we will determine later.
To this end, let us consider $(\tga, \tesa) \in \cB_{\tilde{r}\av}(0_{\nz})$ and proceed by induction.
Let $(\tgap,\tesap) \in \R^{m \times n} \times \dpr$ such that $\col(\tgap,\tesap) := \col(\tga,\tesa) + \ssz f\av(\col(\tga,\tesa))$.
Then, the increment of $\Vav$ along trajectories of~\eqref{eq:algo_closed_loop_average} is given by
\begin{align}
	&	\Delta \Vav(\tga,\tesa)
	:= \Vav(\tgap ,\tesap) - \Vav(\tga,\tesa)
	\notag\\
	& = 
	\! \mav ( \J(\tgap  \! + \! \gstar,\pr) 
	\! - \! \J(\tga \! + \! \gstar,\pr))
	\notag
	\! - \! \ssz\! \left(1 \! - \! \tfrac{\ssz}{2}\right)\!\|\tesa\|^2
	\notag
	\\
	&\stackrel{(a)}{=} 
	\mav\J(\tga +\gstar - \ssz \G(\tga + \gstar,\tesa + \pr),\pr)
	\notag\\
	&\hspace{.4cm}
	-\mav\J(\tga + \gstar - \ssz \G(\tga + \gstar,\pr),\pr)
	\notag\\
	&\hspace{.4cm}
	+\mav\J(\tga + \gstar - \ssz \G(\tga + \gstar,\pr),\pr)
	\notag\\
	&\hspace{.4cm}
	-\mav\J(\tga + \gstar,\pr)
	- \ssz\left(1 - \ssz/2\right)\|\tesa\|^2,\label{eq:pre_deltaVav_intermediate}
\end{align}
where in $(a)$ we add $\pm \mav\J(\tga + \gstar - \ssz \G(\tga+\gstar ,\pr),\pr)$.
By using $\lipp_0$ (cf.~\eqref{eq:lipp0}), we note that 
\begin{align}
	\snorm{\tga - \ssz \G(\tga + \gstar,\tesa + \pr)} \leq \tilde{r}\av + \ssz \lipp_0,\label{eq:maximum_norm}
\end{align}
for all $(\tga,\tesa) \in \cB_{\tilde{r}\av}(0_{\nz})$.
Hence, the bound~\eqref{eq:maximum_norm} ensures $(\tga - \ssz \G(\tga + \gstar,\tesa + \pr),\tesa) \in \cB_{\rmax}(0_{\nz})$ for all $(\tga,\tesa) \in \cB_{\tilde{r}\av}(0_{\nz})$ and $\ssz \in (0,\bar{\ssz}_0\av)$, where $\bar{\ssz}_0\av := (\rmax - \tilde{r}\av)/\lipp_0$
Then, let $\lipp_3, \lipp_4 \!>\! 0$ be defined as
\begin{subequations}\label{eq:lipp_tilde}
	\begin{align}
		\lipp_3  &:=  \max_{(\tga,\tesa) \in \cB_{\rmax}(0_{\nz})}\left\{ \norm{\tfrac{\partial \G(\tga  +   \gstar,\tesa   +   \pr)}{\partial \tga}}\right\}
		\label{eq:lipp3}
		\\
		\lipp_4  &:=  \max_{(\tga,\tesa) \in \cB_{\rmax}(0_{\nz})}\left\{ \norm{\tfrac{\partial \G(\tga  +  \gstar, \tesa  +  \pr)}{\partial \tesa}}\right\}.
		\label{eq:lipp4}
	\end{align}
\end{subequations}
By using~\eqref{eq:lipp3} and expanding $\J$ about $(\tgat+\gstar,\pr)$ at $(\tgat + \gstar - \ssz \G(\tgat + \gstar,\pr),\pr)$, we bound~\eqref{eq:pre_deltaVav_intermediate} as
\begin{align}
	\Delta \Vav(\tga,\tesa)
	& \leq 
	\mav\J(\tga  + \gstar  -  \ssz \G(\tga  +  \gstar,\tesa  +  \pr),\pr)
	\notag\\
	&\hspace{.4cm}
	-\mav\J(\tga +\gstar- \ssz \G(\tga + \gstar,\pr),\pr)
		\label{eq:deltaVav_intermediate}\\
	&\hspace{.4cm}
	-\ssz\mav(1 - \ssz\tfrac{\lipp_3}{2})
	\|\G(\tga + \gstar,\pr)\|^2 
	\notag
	 -  \ssz\left(1 - \tfrac{\ssz}{2}\right)\|\tesa\|^2.
\end{align}
Now, we handle the first two terms in~\eqref{eq:deltaVav_intermediate}.
We expand $\J$ about $(\tga+\gstar,\pr)$ evaluated at $(\tga + \gstar - \ssz \G(\tga +\gstar,\tesa + \pr),\pr)$ and $(\tga +\gstar- \ssz \G(\tga + \gstar,\pr),\pr)$, use~\eqref{eq:lipp3}, the Cauchy-Schwarz inequality, and get
\begin{align}
	&\J(\tga +\gstar- \ssz \G(\tga + \gstar,\tesa + \pr),\pr) 
	\notag\\
	&\hspace{.4cm}
	- \J(\tga +\gstar- \ssz \G(\tga + \gstar,\pr),\pr)
	\notag
	\\
	&\leq 
	\ssz\|\G(\tga+\gstar,\pr)\|
	\|\G(\tga  + \gstar,\tesa + \pr)  - \G(\tga  + \gstar,\pr)\|
	\notag\\
	&\hspace{.4cm}
	 +  \tfrac{\ssz^2\lipp_3}{2}(\|\G(\tga   +  \gstar,\tesa  +  \pr)\|^2
	 +  \|\G(\tga  +  \gstar,\pr)\|^2)
	\notag
	\\
	&\stackrel{(a)}{\leq}
	\ssz\lipp_4\|\G(\tga+\gstar,\pr)\|\|\tesa\|
	\notag\\
	&\hspace{.4cm}
	+\tfrac{\ssz^2\lipp_3}{2}\|\G(\tga +  \gstar,\tesa + \pr) \pm \G(\tga + \gstar,\pr)\|^2
	\notag\\
	&\hspace{.4cm}
	+\tfrac{\ssz^2\lipp_3}{2}\|\G(\tga + \gstar,\pr)\|^2
	\notag
	\\
	&\stackrel{(b)}{\leq}
	\ssz\lipp_4\|\G(\tga+\gstar,\pr)\|\|\tesa\|
	+\ssz^2\lipp_3\lipp_4\|\tesa\|^2
	+ \ssz^2\lipp_3\|\G(\tga +\gstar,\pr)\|^2 
	\notag\\
	&\hspace{.4cm}
	 +  \tfrac{\ssz^2\lipp_3}{2}\|\G(\tga +\gstar,\pr)\|^2,\label{eq:deltaJ}
\end{align}
\vspace{-.03cm}
where in $(a)$ we use~\eqref{eq:lipp4} and add $\pm \G(\tga + \gstar,\pr)$ inside the norm of the second term, 
while $(b)$ uses~\eqref{eq:lipp4} and a standard property of $\norm{\cdot}$\footnote{Given any $v_1, v_2 \! \in\! \R^{n}$, it holds $\norm{v_1 \! - \! v_2}^2 \! \leq\! 2\norm{v_1}^2 \! + \! 2\norm{v_2}^2$.}.
By plugging the bound in~\eqref{eq:deltaJ} into~\eqref{eq:deltaVav_intermediate}, %
we get
\begin{align}
\Delta\Vav(\tga,\tesa) 
&\leq -\ssz\mav(1 - \ssz\tfrac{3\lipp_3}{2})\|\G(\tga + \gstar,\pr)\|^2 
\notag\\
&\hspace{.4cm}
+\ssz\mav\lipp_4\|\G(\tga + \gstar,\pr)\|\|\tesa\|
\notag\\
&\hspace{.4cm}- \ssz(1 - \ssz\tfrac{1 + \mav\lipp_3\lipp_4^2}{2})\|\tesa\|^2,\label{eq:deltaVav_5}
\end{align}
Let us arbitrarily choose $\nu_1,\nu_2 \in (0,1)$.
Then, for all $\ssz \in (0,\bar{\ssz}\av)$ with $\bar{\ssz}\av := \min\{\bar{\ssz}_0\av,\tfrac{2\nu_1}{3\lipp_3},2\nu_2\}$, we further bound~\eqref{eq:deltaVav_5} as
\begin{align}
	\Delta\Vav(\tgat,\tesat) 
	&
	\leq -\ssz\mav\nu_1\|\G(\tga + \gstar,\pr)\|^2 
	\notag\\
	&\hspace{.4cm}
	+\ssz\mav\lipp_4 \|\G(\tga + \gstar,\pr)\|\|\tesa\|
	- \ssz(\nu_2 - \mav\lipp_3\lipp_4^2)\|\tesa\|^2
	\notag
	\\
	&
	\stackrel{(a)}{=} 
	-\ssz\begin{bmatrix}
		\|\G(\tga  +  \gstar,\pr)\|
		\\
		\|\tesa\|
	\end{bmatrix}\T  U(\mav) 
	\begin{bmatrix}
		\|\G(\tga  +  \gstar,\pr)\|
		\\
		\|\tesa\|
	\end{bmatrix}
	,\label{eq:deltaVav_6}
\end{align}
where in $(a)$ we introduce the matrix
\begin{align*}
	U(\mav)  := 
	\begin{bmatrix}
		\mav\nu_1& -\tfrac{\mav\lipp_4}{2}
		\\
		-\tfrac{\mav\lipp_4}{2}& \nu_2 - \mav\lipp_3\lipp_4^2
	\end{bmatrix}.
\end{align*}
We set $\mav \in (0,4\nu_1\nu_2/\lipp_4^2(1+4\nu_1\lipp_3))$.
By Sylvester criterion this implies $U(\mav) > 0$. %
Then, by denoting with $\eta > 0$ the smallest eigenvalue of $U(\mav)$, we bound~\eqref{eq:deltaVav_6} as
\begin{align}
	\Delta\Vav(\tga,\tesa) 
	&\leq - \ssz\eta(\|\G(\tga + \gstar,\pr)\|^2 + \|\tesa\|^2)
	\notag\\
	&\stackrel{(a)}{\leq}
	- \ssz\tfrac{\eta}{\gd}(\J(\tga  +  \gstar,\pr)  - \J(\gstar,\pr)) 
	- \ssz\eta\|\tesa\|^2
	\notag\\
	&\stackrel{(b)}{\leq}
	- \ssz\eta\min\{1/\gd\mav,1\}\Vav(\tga,\tesa),
	\label{eq:deltaVav_4}
\end{align}
where $(a)$ uses~\eqref{eq:gradient_dominance} and $(b)$ uses~\eqref{eq:Vav}.
Moreover, by~\cite[Lemma~3.8]{bu2020lqr} and the Lipschitz continuity of $\G$ in $\cB_{r\av}(0_{\nz})$ (see~\eqref{eq:lipp1}), there exist $c_2 \ge c_1 > 0$ such that
\begin{align*}
	c_1\|\tga\|^2 \leq \J(\tga + \gstar,\pr) - \J(\gstar,\pr) \leq c_2\|\tga\|^2,
\end{align*}
for all $(\tesa,\tga) \in \cB_{\tilde{r}\av}(0_{\nz})$, which, combined with the definition of $\Vav$ (cf.~\eqref{eq:Vav}), implies
\begin{align}
	c_3\|(\tga,\tesa)\|^2 \leq\Vav(\tga,\tesa) \leq c_4\|(\tga,\tesa)\|^2,\label{eq:quadratic_bounds}
\end{align}
for all $(\tesa,\tga) \in \cB_{\tilde{r}\av}(0_{\nz})$, where $c_3 := \min\{c_1\mav,1/2\}$ and $c_4 := \max\{c_2\mav,1/2\}$.
Then, by monotonicity of $\Vav$ (cf.~\eqref{eq:deltaVav_4}) and its quadratic bounds (cf.~\eqref{eq:quadratic_bounds}), we get 
\begin{align}\label{eq:last}
	\tz\ud{+} &\in \cB_{\sqrt{c_4/c_3}\norm{\tz}}(0_{\nz}),
\end{align}
for all $\tz \in \cB_{\tilde{r}\av}(0_{\nz})$, where $\tz\ud{+}$ is the update along~\eqref{eq:algo_closed_loop_average}.
By observing~\eqref{eq:last} and setting $r\av := \sqrt{c_3/c_4}\tilde{r}\av$, we get invariance of $\cB_{r\av}(0_{\nz})$ for system~\eqref{eq:algo_closed_loop_average}.
Once this invariance is achieved, we use~\eqref{eq:deltaVav_4} and~\eqref{eq:quadratic_bounds} to claim exponential stability of the origin for system~\eqref{eq:algo_closed_loop_average} (cf.~\cite[Theorem~13.2]{haddad2011nonlinear}) and the proof follows.

\end{document}